\documentclass[10pt,final]{iopart}
\usepackage{iopams,graphicx,texdraw,epsf,bm}
\def\twofone{\mathop{_{2}{F}_{1}}\nolimits}
\begin{document}
\paper[Boundary critical behaviour at Lifshitz points]{Boundary
  critical behaviour at $\bm{m}$-axial Lifshitz points: the special
  transition for the case of a surface plane parallel to the
  modulation axes}
\author[H.~W. Diehl and S. Rutkevich]{H.~W. Diehl and S.\ Rutkevich%
\footnote{On leave of absence from Institute of Solid State and Semiconductor
Physics, 220072 Minsk, Belarus.}}
\address{Fachbereich Physik, Universit{\"a}t
Duisburg-Essen, Campus Essen, D-45117 Essen, Federal Republic of Germany}
\begin{abstract}
  The critical behaviour of $d$-dimensional semi-infinite systems with
  $n$-component order parameter $\bm{\phi}$ is studied at an $m$-axial
  bulk Lifshitz point whose wave-vector instability is isotropic in an
  $m$-dimensional subspace of $\mathbb{R}^d$. Field-theoretic
  renormalization group methods are utilized to examine the special
  surface transition in the case where the $m$ potential modulation
  axes, with $0\leq m\leq d-1$, are parallel to the surface. The
  resulting scaling laws for the surface critical indices are given.
  The surface critical exponent $\eta_\|^{\rm sp}$, the surface
  crossover exponent $\Phi$ and related ones are determined to first
  order in $\epsilon=4+\case{m}{2}-d$. Unlike the bulk critical
  exponents and the surface critical exponents of the ordinary
  transition, $\Phi$ is $m$~dependent already at first order in
  $\epsilon$. The $\Or(\epsilon)$ term of $\eta_\|^{\rm sp}$ is found
  to vanish, which implies that the difference of $\beta_1^{\rm sp}$
  and the bulk exponent $\beta$ is of order $\epsilon^2$.
\end{abstract}
\submitto{\JPA}

\section{Introduction}\label{sec:intro}

As is well known, systems in critical or near-critical states are
sensitive to the presence of boundaries such as surfaces or walls: the
long-distance behaviour of their local densities and correlation
functions gets modified near boundaries. During the past decades
impressive progress has been made in our understanding of such
boundary critical phenomena
(Binder 1983, Diehl 1986, Cardy 1987, Pleimling 2004).  A prototype
class of systems from whose study considerable insight into general
aspects of boundary critical phenomena has emerged is provided by the
semi-infinite $n$-vector $\phi^4$ models. At bulk criticality, these
models exhibit a variety of physically distinct continuous surface
transitions in sufficiently high space dimensions $d$, called
ordinary, special, extraordinary, and normal. A simplifying feature
they have is that the scale invariance they display at criticality is
of \emph{isotropic} nature: distances along arbitrary directions must
be scaled with the same power of the length-rescaling factor $\ell$.
Furthermore, the correlation lengths $\xi(\bm{e})$ defined through the
exponential decay $\sim \exp(-R/\xi(\bm{e}))$ of the pair correlation
function $\langle\phi(\bm{x})\,\phi(\bm{x}+R\bm{e})\rangle$ along a
given direction specified by a unit vector $\bm{e}$ all diverge
$\sim|T-T_{b,c}|^{-\nu}$ with the same critical index $\nu$, as the
critical bulk temperature $T=T_{c,b}$ is approached via a phase with
unbroken $O(n)$ symmetry. This leaves little room for dependence of
boundary critical behaviour on the orientation of the surface with
respect to crystal axes. We are aware of only one established scenario
for such orientation dependence
(Schmid 1993, Diehl \etal 1997, Leidl and Diehl 1998, Leidl \etal
1998): bcc binary alloys and antiferromagnets in the presence of weak
magnetic fields may map upon coarse graining onto semi-infinite
$\phi^4$ models with or without ordering boundary fields $h_1$,
depending on whether their surface's orientation breaks the symmetry
$\phi\to-\phi$ or not.  If so, they may belong to basins of attraction
of distinct fixed points, such as the normal and ordinary one, so that
their boundary critical behaviour is different.

In the present paper we shall be concerned with boundary critical
behaviour at $m$-axial bulk \emph{Lifshitz points} (LP)
(Hornreich 1980, Selke 1992, Diehl 2002). An LP is a multi-critical
point at which a disordered, a homogeneous ordered and a modulated
ordered phase meet.  A characteristic feature of the scale invariance
that systems exhibit at such points is its \emph{anisotropic} nature:
the coordinates $x_\alpha$, $\alpha =1,\ldots,m$, along any of the $m$
potential modulation axes scale as a nontrivial power
$(x_\beta)^\theta$ of the $d-m$ remaining orthogonal ones, $x_\beta$,
$\beta =m+1,\ldots,d$.  This entails that the orientation of the
surface with respect to the modulation axes matters. Two principal
surface orientations, parallel and perpendicular, can be distinguished
for which the surface normal is directed along an $\alpha$ or $\beta$
axis, respectively.  The boundary critical behaviour that occurs at a
given type of surface transition (ordinary, special, etc) for either
parallel or else perpendicular surface orientation is expected to
correspond to two \emph{distinct} universality classes. Results
obtained via mean-field theory
(Gumbs 1986, Binder and Frisch 1991, Frisch \etal 2000) and Monte
Carlo simulations
(Pleimling 2002) for the ordinary and special transitions of ANNNI
models with free surfaces of both types of orientation are in
conformity with this expectation.

These findings indicate that systems exhibiting anisotropic scale
invariance have potentially richer boundary critical behaviour. Since
systems of this kind are abundant in nature---anisotropic scale
invariant states exist both in and out of equilibrium---a careful
understanding of their boundary critical behaviour is certainly of
interest.

Focusing on the case of parallel surface orientation, we have recently
introduced an appropriate semi-infinite extension of a standard
$n$-vector $\phi^4$ model describing universality classes of
\emph{bulk} critical behaviour at $m$-axial LP points
(Diehl \etal 2003a, 2003b). We argued that this extension represents
the \emph{surface} universality classes of the ordinary, special and
extraordinary transitions for this kind of surface orientation
(Diehl 1986, 1997). A renormalization group (RG) analysis for
dimensions $d=d^*(m)-\epsilon$ below the upper critical dimension
$d^*(m)=4+m/2$ lent support to this claim, giving fixed points we
could identify as describing the corresponding surface universality
classes. However, this identification was tentative: since of all,
only the ordinary transition was analysed in some detail; second, this
was done by relying on the asymptotic validity of the Dirichlet
boundary condition, bypassing thereby the need to determine the
precise location of the ordinary fixed point in the space of surface
interaction constants.

The purpose of the present paper is to complement our previous work
(Diehl \etal 2003a, 2003b) by a detailed analysis of the
\emph{special} transition. The asymptotic (Robin-type) boundary
condition that the theory satisfies at this transition involves the
fixed-point value $\lambda_+^*$ of a dimensionless surface variable
$\lambda$, a number which depends on the order of the loop expansion.
A trick analogous to the one employed in the analysis of the ordinary
transition by which the computation of $\lambda_+^*$ can be avoided
does not exist. This entails a \emph{qualitative} difference between
the analysis of the $m>0$ special transition described below and that
of its $m=0$ counterpart at bulk \emph{critical} points
(Diehl and Dietrich 1981, 1983): in the latter case, the variable
$\lambda$ is missing, and the fixed-point value of the sole remaining
renormalized surface interaction constant (the surface enhancement
$c$) is zero to all orders in perturbation theory. Unlike the $m>0$
case, projection onto the hyper-plane of critical renormalized surface
variables is trivial.

We shall determine the surface correlation index $\eta_\|^{\rm sp}$
and the surface crossover exponent $\Phi$ to first order in
$\epsilon$. Our RG analysis reveals that below the upper critical
dimension, all other surface critical exponents of the special
transition can be expressed in terms of these two independent surface
critical exponents, along with bulk critical indices. While this is
just as in the $m=0$ case of the special transition at bulk critical
points
(Diehl and Dietrich 1981, 1983, Diehl 1986), our result that the
$\Or(\epsilon)$ contribution to $\eta_\|^{\rm sp}$ vanishes if $m>0$
(though not for $m=0$) is a remarkable difference.  Another unusual
feature of our results is that the crossover exponent $\Phi$ starts to
depend on $m$ already at \emph{linear} order in $\epsilon$. By
contrast, the $\epsilon$-expansions of both the bulk critical
exponents as well as the surface critical exponents of the ordinary
transition display first $m$-dependent deviations from their $m=0$
analogues only at quadratic order in $\epsilon$.

The remainder of this paper is organized as follows. In the next
section, we define our model and briefly recapitulate the relevant
background knowledge required in what follows.
Section~\ref{sec:epsexp} deals with the RG analysis of the special
transition. We derive the scaling forms of the multi-point cumulants
involving order parameter fields $\bm{\phi}$ on and off the boundary,
give some of the implied scaling laws, and present our
$\epsilon$-expansion results.  The latter are utilized to estimate the
values of the surface critical exponents $\eta_\|^{\rm sp}$ and
$\beta_1^{\rm sp}$, as well as the surface crossover exponent for the
uniaxial scalar case in three dimensions, i.e.\ the ANNNI model. In
section~\ref{sec:RGipt}, RG~improved perturbation theory is employed
to investigate the surface susceptibility $\chi_{11}$. Making an
explicit one-loop calculation, we verify the predicted scaling
behaviour, corroborating thereby our identification of the special
transition. A short discussion and concluding remarks follow in the
final section. Finally, there are two appendices explaining
computational details.

\section{Model and background}
\label{sec:model}

Following
Diehl \etal (2003a, 2003b), we consider a model defined by the
Hamiltonian
\begin{equation}
  \label{eq:Hamf}
    {\mathcal{H}}={\int_{\mathfrak{V}}}\mathcal{L}_{\rm b}(\bm{x})\,
    \rmd V+
    {\int_{\mathfrak{B}}}\mathcal{L}_1(\bm{x})\,\rmd A\;,
\end{equation}
with the bulk density
\begin{equation}
  \label{eq:Lb}\fl
 {\mathcal L}_{\rm b}(\bm{x})=\frac{\mathring{\sigma}}{2}\,
  \bigg(\sum_{\alpha=1}^m\partial_\alpha^2\bm{\phi} \bigg)^2
    +\frac{1}{2}\,
\sum_{\beta=m+1}^d{(\partial_\beta\bm{\phi})}^2
+\frac{\mathring{\rho}}{2}\,
\sum_{\alpha=1}^m{(\partial_\alpha\bm{\phi})}^2
 +\frac{\mathring{\tau}}{2}\,
\bm{\phi}^2+\frac{\mathring{u}}{4!}\,|\bm{\phi} |^4
\end{equation}
and the surface density
\begin{equation}
  \label{eq:L1}
  \mathcal{L}_1(\bm{x})=\frac{\mathring{c}}{2}\,\bm{\phi}^2 +
\frac{\mathring{\lambda}}{2}\,
\sum_{\alpha=1}^m{(\partial_\alpha\bm{\phi})}^2 \;,
\end{equation}
where the volume and surface integrals $\int_{\mathfrak{V}}\rmd V$ and
$\int_{\mathfrak{B}}\rmd A$ extend over the half-space
${\mathfrak{V}=\{\bm{x}=(\bm{r},z)\mid\bm{r}\in\mathbb{R}^{d-1},0\leq
  z<\infty\}}$ and the boundary plane
${\mathfrak{B}=\{(\bm{r},0)\mid\bm{r}\in\mathbb{R}^{d-1}\}}$ (`the
surface'), respectively, and the notation $\partial_\alpha
=\partial/\partial x_\alpha $ and $\partial_\beta =\partial/\partial
x_\beta $, with $1\leq \alpha \leq m$ and $m<\beta \leq d$, is used.
The order parameter is an $n$-vector
$\bm{\phi}=(\phi_a,a=1,\ldots,n)$.

Let us introduce the $(N+M)$-point cumulants involving $N$ fields
$\phi_{a_i}(\bm{x}_i)$ off the surface and $M$ boundary fields
$\phi_{b_j}^{\mathfrak{B}}(\bm{r}_j)\equiv \phi_{b_j}(\bm{r}_j,z=0)$:
\begin{equation}
  \label{eq:GNM}
  G^{(N,M)}(\bm{x};\bm{r})\equiv\bigg\langle
  \bigg[\prod_{i=1}^N\phi_{a_i}(\bm{x}_i)\bigg]
    \bigg[\prod_{j=1}^M \phi_{b_j}^{\mathfrak{B}}(\bm{r}_j)\bigg]
    \bigg\rangle^{\rm cum}\;,
\end{equation}
where $\bm{x}$ and $\bm{r}$ are shorthands for the sets of all points
$\bm{x}_i$ and $\bm{r}_j$ off or on the boundary, respectively.  From
previous work
(Diehl \etal 2003a, 2003b) we know that the ultraviolet (uv)
singularities of these functions can be absorbed for $d\leq d^*(m)$ by
means of the following re-parametrizations:
\begin{eqnarray}
  \label{eq:bulkrep}
  \bm{\phi}=Z_\phi^{1/2}\,\bm{\phi}_{\rm ren}\;,\quad
  \mathring{\sigma}=Z_\sigma\,\sigma\;,
\quad  \mathring{u}\,{\mathring{\sigma}}^{-m/4}\,F_{m,\epsilon}=
   \mu^\epsilon\,Z_u\,u\;,\nonumber\\
\mathring{\tau}-\mathring{\tau}_{\rm LP}=
\mu^2\,Z_\tau{\big[\tau+A_\tau\,\rho^2\big]}\;,
\quad
\left(\mathring{\rho}-\mathring{\rho}_{\rm LP}\right)\,
{\mathring{\sigma}}^{-1/2}=\mu\,Z_\rho\,\rho\;,
\end{eqnarray}
\begin{eqnarray}
  \label{eq:surfrep}
  \bm{\phi}^{\mathfrak{B}}=(Z_\phi
  Z_1)^{1/2}\,\bm{\phi}^{\mathfrak{B}}_{\rm ren}\;,
\quad
\mathring{\lambda}\,\mathring{\sigma}^{-1/2}&=&\lambda+
P_\lambda(u,\lambda,\epsilon)\;,\nonumber\\
 \mathring{c}-\mathring{c}_{\rm sp}=
  \mu\,Z_c{\big[c+A_c(u,\lambda,\epsilon)\,\rho\big]}\;,
\end{eqnarray}
where
\begin{equation}\label{Fmeps}
F_{m,\epsilon}=
\frac{\Gamma{\left(1+{\epsilon/ 2}\right)}
\,\Gamma^2{\left(1-{\epsilon/ 2}\right)}\,
\Gamma{\left({m}/{4}\right)}}{(4\,\pi)^{({8+m-2\,\epsilon})/{4}}\,
\Gamma(2-\epsilon)\,
\Gamma{\left({m}/{2}\right)}}
\end{equation}
is a convenient normalization constant. Further, $\mathring{\tau}_{\rm
  LP}$ and $\mathring{\rho}_{\rm LP}$ are the critical bare values of
the LP; in our renormalization scheme based on dimensional
regularization and the $\epsilon$-expansion, these fluctuation-induced
shifts vanish. The same applies to $\mathring{c}_{\rm sp}$, the
critical value of the bare surface enhancement $\mathring{c}$ at which
the special transition occurs.

The bulk renormalization factors
$Z_{\phi,\tau,\rho,u}=Z_{\phi,\tau,\rho,u}(u,\epsilon)$ are power
series in $u$, the renormalized coupling constant, and Laurent series
in $\epsilon$; since we employ the scheme of minimal subtraction of uv
poles, their regular part at $\epsilon=0$ is exactly 1. The boundary
renormalization factors $Z_{1,c}=Z_{1,c}(u,\lambda,\epsilon)$ have a
similar form, except that the series coefficients depend on $\lambda$.
Finally, $P_\lambda$ and $A_c$ are renormalization functions of the
form
\begin{equation}
  \label{eq:Plambda}
  P_\lambda(u,\lambda,\epsilon)=\sum_{i,j=1}^\infty
  P_{\lambda}^{(i,-j)}(\lambda)\,u^i\,\epsilon^{-j}
  =\sum_{i,j=1}^\infty\sum_{k=0}^\infty P_\lambda^{(i,-j;k)}\,
     u^i\,\epsilon^{-j}\,\lambda^k\;.
\end{equation}

Two important features should be appreciated: first, although the
$\Or(u)$ term of $P(u,\lambda,\epsilon)$ is proportional to $\lambda$,
its $\Or(u^2)$ contribution is not, and hence does \emph{not} vanish
for $\lambda=0$. Thus $P(u,\lambda=0,\epsilon)\neq 0$, i.e.\ a nonzero
value of the bare interaction constant $\mathring{\lambda}$ gets
generated by the $\phi^4$ interaction
(Diehl \etal 2003a). Second, the renormalization of $\mathring{\tau}$
and $\mathring{c}$ mixes the corresponding renormalized quantities
$\tau$ and $c$ with $\rho^2$ and $\rho$, respectively. The
consequences (worked out in
(Diehl \etal (2003a) and below) should be no surprise: the role of the
scaling fields $\tau$ and $c$ of the Gaussian theory is taken over by
combinations of the form
\begin{equation}
  \label{eq:gtaucintro}
  g_\tau(\tau,\rho,u)=\tau+C^\tau_{\rho^2}(u)\,\rho^2\;, \quad
   g_c(c,\rho,u,\lambda)=c+{C}^c_{\rho}(u,\lambda)\,\rho\;.
\end{equation}

To become more specific, it is necessary to recall the RG equations of
the renormalized cumulants $G^{(N,M)}_{\rm ren}=
Z_\phi^{-(N+M)/2}Z_1^{-M/2} \,G^{(N,M)}$. They read
\begin{equation}
  \label{eq:RGE}
 {\bigg[ \mu\partial_\mu+\sum_{\wp
    =u,\sigma,\tau,\rho,c,\lambda}\beta_\wp\,\partial_\wp
  +\frac{N+M}{2}\,\eta_\phi+
      \frac{M}{2}\,\eta_1\bigg]}G^{(N,M)}_{\rm ren}=0\;.
\end{equation}
The beta functions, defined by
$\beta_\wp\equiv\left.\mu\partial_\mu\right|_0\wp$, where
$\partial_\mu|_0$ denotes a derivative at fixed bare interaction
constants, can be expressed in terms of the exponent functions
$\eta_{\wp,\phi,1}\equiv\left.\mu\partial_\mu\right|_0\ln
Z_{\wp,\phi,1}$ and $P_\lambda$ as
\begin{eqnarray}
  \label{eq:betauetc}
  \beta_u(u,\epsilon)&=&-u\,[\epsilon+\eta_u(u)]\;,
\nonumber\\
\beta_\sigma(u,\sigma)&=&-\sigma\,\eta_\sigma(u)\;,
\nonumber\\
\beta_\tau(u,\tau,\rho)&=&-\tau\,[2+\eta_\tau(u)]-\rho^2\,b_\tau(u) \;,
\\
\beta_\rho(u,\rho)&=&-\rho\,[1+\eta_\rho(u)]\;,
\nonumber\\
\beta_c(u,\lambda,\rho,c)&=&
-c\,[1+\eta_c(u,\lambda)]-\rho\,b_c(u,\lambda)\nonumber\;,
\end{eqnarray}
and
\begin{equation}
  \label{eq:betalambda}
  \beta_\lambda(u,\lambda)=\frac{-\beta_u(u,\epsilon)\,\partial_u
    P_\lambda(u,\lambda,\epsilon)}{1+\partial_\lambda
    P_\lambda(u,\lambda,\epsilon)}\;,
\end{equation}
where the contributions involving $b_\tau$ and $b_c$ are produced by
the terms proportional to $A_\tau$ and $A_c$ in
equations~(\ref{eq:bulkrep}) and (\ref{eq:surfrep}), respectively.

Explicit two-loop results for the bulk quantities
$Z_{\phi,\sigma,\tau,u}$, $\beta_u$, and
$\eta_{\phi,\sigma,\tau,\rho}$ can be found in
Shpot and Diehl (2001); see also
Diehl and Shpot (2000). The bulk function $A_\tau$ is given to order
$u$ in equation (17) of
Diehl \etal (2003a), where also the renormalization functions $Z_1$,
$Z_c$, and $P_\lambda$ of the semi-infinite system were obtained to
the same order in $u$ for general values of $\lambda\geq 0$; see its
equations (21)--(23). The remaining surface counter-term $A_c$ is
computed in \ref{app:Ac}; we obtain
\begin{equation}
  \label{eq:Ac}\fl
  A_c(u,\lambda,\epsilon)=\frac{n+2}{3}\,\frac{u}{2\epsilon}
  \,{\left[\frac{1-i_1(\lambda;m)-\lambda\,i_1'(\lambda;m)}{2\lambda}
 - \frac{\sqrt{\pi} \,\Gamma[(m+2)/4)]}{2\,\Gamma({m}/{4})}\right]}+
\Or(u^2)\;.
\end{equation}

Utilizing these results gives the beta function
\begin{equation}
  \label{eq:betalambdares}
  \beta_\lambda(u,\lambda)= -\frac{n+2}{6}\,i_1(\lambda;m)\,\lambda\,u
   +2u^2\,P_\lambda^{(2,-1)}(\lambda)+ \Or(u^3)
\end{equation}
and the exponent functions
\begin{equation}
  \label{eq:eta1res}
  \eta_1(u,\lambda)= -\frac{n+2}{6}\,i_1(\lambda;m)\,u
   + \Or(u^2)
\end{equation}
and
\begin{equation}
  \label{eq:etacres}
  \eta_c(u,\lambda)= -\frac{n+2}{6}\,
  [1+\lambda\, i'_1(\lambda;m)]\,u
   + \Or(u^2)\;.
\end{equation}
Here $i_1(\lambda;m)$ and $i'_1(\lambda;m)$ are integrals defined
through
\begin{eqnarray}
  \label{eq:i1}
  i_1(\lambda;m)&\equiv& i_1(\lambda,\epsilon=0;m)\;, \\
i_1'(\lambda;m)&\equiv& i'_1(\lambda,\epsilon=0;m)\;, \\
i_1'(\lambda,\epsilon;m)&\equiv&\partial i_1(\lambda,\epsilon;m)/
\partial\lambda\;,\\
i_1(\lambda,\epsilon;m)&\equiv&
\left\langle\frac{1-\lambda\,t}{1+\lambda\,t}
\right\rangle_{\epsilon,m}\;,
\end{eqnarray}
where we have introduced the convenient notation $\langle
f(t)\rangle_{\epsilon,m}$ for the normalized average
\begin{equation}
  \label{eq:aveps}
  \langle f(t)\rangle_{\epsilon,m}\equiv  \frac{1}{{\mathcal
      N}_{\epsilon,m}}\,\int_0^1\rmd 
    t\,f(t)\,t^{(m-2)/2}\,(1-t^2)^{(2-2\epsilon-m)/4} 
\end{equation} 
over $t$. The normalization factor is given by
\begin{equation}
\label{eq:normN}\fl 
{\mathcal N}_{\epsilon,m}\equiv \int_0^1\rmd 
    t\,t^{(m-2)/2}\,(1-t^2)^{(2-2\epsilon-m)/4}= 
    \case12\,B[m/4,(6-2\epsilon-m)/4] 
\end{equation} 
where $B(a,b)$ is the Euler beta function.  This choice of ${\mathcal
  N}_{\epsilon,m}$ ensures that
\begin{equation} 
  \label{eq:i1lam1} 
  i_1(0,\epsilon;m)=1\,. 
\end{equation}

The $t$-integral required for $i_1(\lambda;m)$ converges for
${0<m<6}$.  Since the normalization factor varies as $1/{\mathcal
  N}_{0,m}\sim (6-m)$ for $m\to 6$, the integral's singularity
$\sim(6-m)^{-1}$ gets cancelled to produce the finite limiting value
\begin{equation} 
  \label{eq:i1six} 
  i_1(\lambda;6-)\equiv \lim_{m\to 6-}i_1(\lambda;m)= 
  \frac{1-\lambda}{1+\lambda}\;. 
\end{equation} 
 
For general values $m\in(0,6)$, the integral $i_1(\lambda;m)$ can be 
expressed in terms of hyper-geometric functions $\twofone$ and 
elementary ones. Term-wise integration of the integrand's Taylor 
series in $\lambda$ or evaluation via {\sc Mathematica}%
\footnote{{\sc Mathematica}, version 4.1, a product of Wolfram Research.} 
leads to 
\begin{equation}\fl 
  \label{eq:i1lamres} 
  i_1(\lambda;m)= 2\;{\twofone}{\left(1,\case{m}{4}; 
    \case{3}{2};{\lambda }^2\right)}-1 + 
    \,\frac{\sqrt{\pi}\,\Gamma[(m-2)/4]}{\Gamma (m/4)\,\lambda}\, 
    {\left[1 - {\left( 1 - {\lambda }^2 \right)}^{(2 - 
          m)/4}\right]}\;, 
\end{equation} 
which simplifies considerably if $m=2$ or $m=6$, giving 
\begin{equation} 
  \label{eq:i1lamm2} 
  i_1(\lambda;2)=2\,\lambda^{-1}\,\ln(1+\lambda)-1 
\end{equation} 
and 
\begin{equation} 
  \label{eq:i1lamm4} 
  i_1(\lambda;4)=\frac{\pi}{\lambda} - 
    \frac{2\,\arccos (\lambda )} 
     {\lambda\,\sqrt{1 - \lambda^2}}-1\;, 
\end{equation} 
respectively.  Explicit plots of the functions 
$-\lambda\,i_1(\lambda;m)$ for $m=1,2,\ldots,6$ are displayed in 
figure 1 of  
Diehl \etal (2003a). 
 
\begin{figure}[bth] 
  \centering \includegraphics[width=20em]{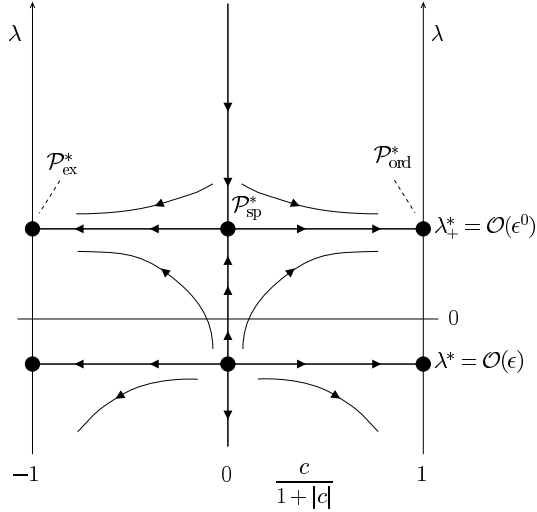} 
  \caption{Schematic picture of the RG flow in the $c\lambda$~plane at 
    $\rho=\tau=0$ and $u=u^*$, showing the fixed points 
    ${\mathcal{P}}_{\rm ord}^*$, ${\mathcal{P}}_{\rm sp}^*$, and 
    ${\mathcal{P}}_{\rm ex}^*$ specified in equation (\ref{eq:Pfix}).} 
  \label{fig:flow} 
\end{figure}

Setting the relevant bulk variables $\tau$ and $\rho$ to zero and the 
coupling constant $u$ to the nontrivial zero of $\beta_u$ for 
$\epsilon>0$, namely%
\footnote{The expansion of $u^*$ to $\Or(\epsilon^2)$ can be found in 
  equations (60) of  
Shpot and Diehl (2001) or equation (42) of  
Diehl \etal (2003a). It 
  will not be needed in the following because our subsequent analysis is 
  based on one-loop results for the surface RG functions $\beta_c$ and 
  $\eta_{1,c}$.} 
\begin{equation} 
  \label{eq:ustar} 
  u^*=\frac{2\epsilon}{3}\,\frac{9}{n+8}+\Or(\epsilon^2)\;, 
\end{equation} 
we can analyse the RG flow and determine the fixed points in the 
$c\lambda$ plane.  This yields the schematic flow picture shown in 
figure \ref{fig:flow} 
Diehl \etal (2003a). %
 
The fixed points 
\begin{eqnarray} 
  \label{eq:Pfix} 
\begin{array}{llr} 
  {\mathcal{P}}_{\rm ord}^*:& 
   \quad(c^*_{\rm ord}=\infty,&\lambda=\lambda^*_+) \;,\\[0.5em] 
  {\mathcal{P}}_{\rm sp}^*:& 
   \quad(c^*_{\rm sp}=0,&\lambda=\lambda^*_+)\;,\\[0.5em] 
  {\mathcal{P}}_{\rm ex}^*:& 
   \quad(c^*_{\rm ex}=-\infty,&\lambda=\lambda^*_+)\;,\end{array} 
\end{eqnarray} 
located at the positive value 
\begin{equation} 
  \label{eq:lambdaplusstar} 
  \lambda_+^*(m)=\lambda_0(m)+ 
  \frac{72\,P_\lambda^{(2,-1)}(\lambda_0)\,\epsilon}{(n+2)(n+8)\,%
    \lambda_0\, i_1'(\lambda_0;m)}  + O(\epsilon^2)\;, 
\end{equation} 
should describe the ordinary, special and extraordinary transitions, 
respectively. Here $\lambda_0=\lambda_0(m)$ are the real positive 
zeros of the functions $i_1(\lambda;m)$; for later use we list their 
values in table~\ref{Tab:zeros}. 
\begin{table}[bth] 
\caption{\label{Tab:zeros}Zeros $\lambda_0(m)$, derivatives 
  $i'_1(\lambda;m)$ and coefficients $1+\lambda_0 i_1'(\lambda_0;m)$ 
  appearing in the $\Or(\epsilon)$ terms of $\eta_c^*$; see equation 
  (\ref{eq:Phiovnu}). In the second column the exact limiting values 
  for $m\to 0$ are listed. The values for $m=1,\ldots,5$ are approximate 
  numerical numbers; those for $m=6$ are exact.} 
\begin{tabular}{@{}llllllll} 
\br 
$m$&0&1 &2&3&4&5&6\\ 
\mr\lineup 
$\lambda_0(m)$&$\infty$&6.20921&2.51286&1.70176&1.34277&1.13629&1\\ 
\mr 
$-i_1'[\lambda_0(m);m]$&0&0.05505&0.17138&0.27332&0.35949&0.43400& 
$\case12$\\ 
\mr 
$1+\lambda_0(m)\, i_1'[\lambda_0(m);m]$&1&0.65816&0.56934&0.53488& 
0.51729&0.50685&$\case12$\\ 
\br 
\end{tabular} 
\end{table} %
We have included the limiting values for $m\to 0$. Of course, for 
$m=0$, the variable $\lambda$ becomes meaningless. However, considering 
$m$ as a continuous variable, we can ask what happens to the RG flow, 
the zero $\lambda_0(m)$ and the critical exponents as $m\to 0$. 
Utilizing either the integral representation (\ref{eq:i1}) of 
$i_1(\lambda;m)$ or the explicit form (\ref{eq:i1lamres}) one easily 
verifies that $\lim_{m\to 0}i_1(\lambda;m)=1$ for arbitrary fixed 
$\lambda\geq 0$. Hence the root $\lambda_0(m)$ must approach infinity 
in this limit. To obtain its limiting behaviour, one can work out the 
asymptotic expansion of $i_1$ for $\lambda\to \infty$ at fixed $m<2$. 
A straightforward analysis yields 
\begin{equation} 
  \label{eq:i1as} 
  i_1(\lambda;m)\mathop{=}_{\lambda\to\infty}-1+\lambda^{-m/2}\, 
  \frac{2\pi^{3/2}\csc(m\pi/2)}{\Gamma(m/4]\Gamma[6-m)/4]} 
  +O(\lambda^{-1})\;, 
\end{equation} 
which implies the small-$m$ form 
\begin{equation} 
  \label{eq:lambdazeroas} 
  \lambda_0(m)\mathop{\approx}_{m\to 0}2^{(2-m)/m}\,\e\;. 
\end{equation} 
An evident consequence of these two equations is that the product 
$\lambda_0(m)\,i'_1[\lambda_0(m);m]$ (which will be needed below) 
vanishes for $m\to 0$. 
 
\section{RG analysis of the special transition} 
\label{sec:epsexp} 
 
In order to analyse the asymptotic behaviour described by the special 
fixed point ${\mathcal{P}}_{\rm sp}^*$, we must determine the relevant 
bulk and scaling fields that can be constructed from $\tau$, $\rho$, 
and $c$. Following  
Diehl \etal (2003a), we choose the coefficient $C^\tau_{\rho^2}(u)$ in 
equation~(\ref{eq:gtaucintro}) such that the running variable 
$\bar{g}_\tau(\ell)$ into which $g_\tau$ transforms under a scale 
change $\mu\to\mu\ell$ satisfies 
\begin{equation} 
  \label{eq:gtauflow} 
 \ell\frac{d}{d\ell}\, 
 \bar{g}_\tau(\ell)=-[2+\eta_\tau(\bar{u})]\,\bar{g}_\tau(\ell)\;,\quad 
 \bar{g}_\tau(1)=g_\tau\;, 
\end{equation} 
where the asterisk on functions of $u$ means their value at $u=u^*$. 
 
Upon taking into account the flow equations 
\begin{equation} 
  \label{eq:rhoflow} 
   \ell\frac{d}{d\ell}\, 
 \bar{\wp}(\ell)=\beta_\wp[\bar{u}(\ell),\ldots]\,, \quad 
 \bar{\wp}(1)=\wp\;,\quad\wp=u,\sigma,\tau,\rho,c,\lambda\,, 
\end{equation} 
one easily derives the condition 
\begin{equation} 
  \label{eq:ctaurhocond} 
  b_\tau(\bar{u}) =\big[\eta_\tau(\bar{u})-2\eta_\rho(\bar{u}) 
  + \beta_u(\bar{u},\epsilon)\,\partial_{\bar{u}}\big] 
    C^\tau_{\rho^2}(\bar{u})  \;, 
\end{equation} 
which is easily solved for $\bar{u}$ near $u^*$ to obtain
\begin{equation} 
  \label{eq:ctaurhosol} 
  C^\tau_{\rho^2}(\bar{u})=\frac{b_\tau^*}{\eta_\tau^*-2\eta_\rho^*} 
  +\Or(\bar{u}-u^*)\;. 
\end{equation} 
 
We fix the coefficient $C_\rho^c$ of the scaling field $g_c$ in a 
similar fashion. Requiring that 
\begin{equation} 
  \label{eq:gcflow} 
 \ell\frac{d}{d\ell}\, 
 \bar{g}_c(\ell)=-[1+\eta_c(\bar{u},\bar{\lambda})]\, 
 \bar{g}_c(\ell)\;,\quad 
 \bar{g}_c(1)=g_c\;, 
\end{equation} 
we arrive at the condition 
\begin{equation} 
  \label{eq:ccrhocond}\fl 
  b_c(\bar{u},\bar{\lambda}) 
  =\big[\eta_c(\bar{u},\bar{\lambda}) -\eta_\rho(\bar{u}) + 
  \beta_u(\bar{u},\epsilon)\,\partial_{\bar{u}}+ 
  \beta_\lambda(\bar{u},\bar{\lambda})\, \partial_{\bar{\lambda}}\big] 
  C^c_{\rho}(\bar{u},\bar{\lambda})\;, 
\end{equation} 
which in turn yields 
\begin{equation} 
  \label{eq:ccrhosol} 
  C^c_{\rho}(\bar{u},\bar{\lambda})=\frac{b_c^*}{\eta_c^*-\eta_\rho^*} 
  +\Or{\big(\bar{u}-u^*,\bar{\lambda}-\lambda_+^*\big)} 
\end{equation} 
for $(u,\lambda)$ sufficiently close to the fixed-point values 
$(u^*,\lambda_+^*)$, where the asterisk on functions of $u$ and 
$\lambda$ indicates their values for $(u,\lambda)=(u^*,\lambda_+^*)$, 
i.e.\ at the fixed point ${\mathcal P}^*_{\rm sp}$. 
 
Clearly, the  mixing of $\tau$ and $c$ with $\rho$ should vanish as 
$m\to 0$. This is indeed the case as we see from the explicit $\Or(u)$ 
results 
\begin{equation} 
  \label{eq:btaures} 
  b_\tau(u)=\frac{n+2}{3}\,\frac{mu}{16}+\Or(u^2) 
\end{equation} 
and 
\begin{equation} 
  \label{eq:bcres}\fl 
  b_c(u,\lambda) 
=  -\frac{n+2}{3}\,{\left[\frac{1-i_1(\lambda;m) 
    -\lambda\,i_1'(\lambda;m)}{4\lambda}   
 - \frac{\sqrt{\pi} \,\Gamma[(m+2)/4)]}{4\,\Gamma({m}/{4})}\right]}u 
 +\Or(u^2)\;, 
\end{equation} 
the first of which was obtained in  
Diehl \etal (2003a), while the second follows from equation 
(\ref{eq:Ac}). Note that the $\Or(u)$ terms of both $b_\tau$ and $b_c$ 
vanish in the limit $m\to 0$ quite generally, not only at the fixed 
point ${\mathcal P}^*_{\rm sp}$. 
 
From equations (\ref{eq:gtauflow}) and (\ref{eq:gcflow}) we can read 
off the RG eigenexponents of the scaling fields $g_\tau$ and $g_c$. 
Writing them as $1/ \nu $ and $\Phi/ \nu$, respectively, we identify 
the standard correlation-length exponent 
\begin{equation} 
  \label{eq:nudef} 
\nu=(2+\eta_\tau^*)^{-1} 
\end{equation} 
(denoted by $\nu_{l2}$ in  
Shpot and Diehl (2001), Diehl \etal (2003a)) and the surface crossover 
exponent 
\begin{equation} 
  \label{eq:Phi} 
  \Phi=\nu\,(1+\eta_c^*)\;. 
\end{equation} 
 
Let us also introduce the bulk anisotropy exponent 
\begin{equation} 
  \label{eq:thetadef} 
  \theta=\case14(2+\eta_\sigma^*)\;, 
\end{equation} 
the familiar bulk critical exponents 
\begin{equation} 
  \label{eq:etanudef} 
  \eta=\eta_\phi^*\;,\quad 
  \beta=\case12\nu[d-2+\eta+m(\theta-1)]\;, 
\end{equation} 
(denoted by $\eta_{l2}$ and $\beta_l$, respectively, in 
Shpot and Diehl (2001), Diehl \etal (2003a)), the bulk crossover 
exponent 
\begin{equation} 
\label{eq:varphidef} 
  \varphi=\nu(1+\eta_\rho^*) 
\end{equation} 
as well as the surface critical indices 
\begin{equation} 
\label{eq:etapar} 
  \eta_\|^{\rm sp}=\eta+\eta_1^* 
\end{equation} 
and 
\begin{equation} 
\label{eq:beta1} 
  \beta_1^{\rm sp}=\case12\nu[d-2+\eta_\|^{\rm sp}+m(\theta-1)]=\beta 
  +\case12\nu \eta_1^* 
\end{equation} 
of the special transition. 
 
It is now straightforward to solve the RG equation (\ref{eq:RGE}) via 
characteristics and derive the asymptotic scaling forms of the 
cumulants $G^{(N.M)}$ near the special transition. Setting 
$\mu=\sigma=1$ for notational simplicity and neglecting corrections to 
scaling, we obtain 
\begin{eqnarray}\fl 
  \label{eq:GNMscf} 
G^{(N,M)}_{\rm ren}(r_\alpha,r_\beta,z;\rho,g_\tau,u,g_c,\lambda) 
\nonumber\\ \lo{\approx} g_\tau^{N\beta+M\beta_1^{\rm sp}}\, 
{\mathcal G}^{(N,M)}{\big[ r_\alpha 
  g_\tau^{\nu\,\theta},r_\beta g_\tau^{\nu},zg_\tau^{\nu}; \rho 
  g_\tau^{-\varphi},g_cg_\tau^{-\Phi}\big]}\;. 
\end{eqnarray} 
 
The special choice $N=0$ and $M=1$ shows that our identification 
(\ref{eq:beta1}) of $\beta_1^{\rm sp}$ is correct. On the other hand, 
our definition (\ref{eq:etapar}) of $\eta_\|^{\rm sp}$ may need 
explanation. Noting that the surface cumulant $G^{(0,2)}_{\rm 
  ren}(\bm{r},\bm{0})$ should have a well-defined limit as $g_\tau$, 
$\sigma$ and $g_c$ approach zero, we can read off from equation 
(\ref{eq:GNMscf}) that this function decays at the special transition 
as 
\begin{equation}  
  \label{eq:etaparmeaning}  
  G^{(0,2)}_{\rm ren}(\bm{r},\bm{0})\sim  
  \cases{r^{-[d-2+m(\theta-1)+\eta_\|]}&for $r_\alpha=0$,  
    \\ r^{-[d-2+m(\theta-1)+\eta_\|]/\theta}&for $r_\beta =0$}  
\end{equation}  
in the limit of large separations $r$ (parallel to the surface). This  
translates into the small-momentum behaviour  
\begin{equation}  
  \label{eq:surfsusc}  
  \chi_{11}(\bm{p})\sim\cases{%
p^{\eta_\|-1}\,,&for $p_\alpha=0$,\\  
p^{(\eta_\|-1)/ \theta}\,,&for $p_\beta=0$,\\}  
\end{equation}  
of the momentum-dependent local surface susceptibility 
$\chi_{11}(\bm{p})=\hat{G}^{(0,2)}(\bm{p})$ (its Fourier transform), 
where $\bm{p}=(p_\alpha,p_\beta)$ is the wave-vector conjugate to 
$\bm{r}=(r_\alpha,r_\beta)$. We have suppressed the specification 
`${\rm sp}$' on $\eta_\|$ in these two equations because analogous 
results hold in the case of the ordinary transition, with $\eta_\|$ 
given by $\eta_\|^{\rm ord}$, its analogue for the ordinary transition 
(Diehl \etal 2003a, 2003b). For $m=0$, the above 
dependences on $r$ and $p$ for ${r_\alpha=0}$ and ${p_\alpha=0}$, 
respectively, reduce to the familiar power laws for the case of the 
special transition at a bulk \emph{critical} point  
(Diehl and Dietrich  1981, 1983). 
Henceforth we will refer to this surface transition as the 
`critical-point (CP) special transition', calling its $m>0$ 
analogues we are concerned with here `Lifshitz-point (LP) special 
transitions'. 
 
Note also that the scaling forms (\ref{eq:GNMscf}) can be exploited in 
a straightforward fashion to generalize the standard scaling laws of 
the CP case to $m>0$. Consider, for example, the surface critical 
exponent $\eta_\perp$ defined through the decay 
\begin{equation} 
  \label{eq:G11decay} 
  G^{(1,1)}[(\bm{r},z),\bm{r}]\sim z^{-[d-2+m(\theta-1)+\eta_\perp]} 
\end{equation} 
at the LP.%
\footnote{This exponent may be defined more generally 
  through the decay of the pair correlation function 
  $G^{(2,0)}(\bm{x},\bm{x}')$ at criticality as $\bm{x}'$ moves away 
  from the surface along a direction perpendicular (or at least not 
  parallel) to it while $\bm{x}'$ is fixed, similarly as in the CP 
  case  
  (Diehl and Dietrich 1981, 1983, Diehl 1986).}  Upon introducing the 
momentum-dependent layer surface susceptibility 
\begin{equation} 
  \label{eq:chi1} 
  \chi_1(\bm{p})\equiv {\int_0^\infty}{\rmd}z\, 
  \hat{G}^{(1,1)}(\bm{p};z)\;, 
\end{equation} 
we find from equation (\ref{eq:G11decay}) the small-momentum behaviour$\chi_1(\bm{p})\sim p^{\eta_\perp -2}$ or 
${\sim}~p^{(\eta_\perp-2)/\theta}$, depending on whether $p_\alpha=0$ 
or $p_\beta =0$. The result (\ref{eq:GNMscf}) implies that the 
familiar scaling law 
\begin{equation} 
  \label{eq:etaperp} 
  \eta_\perp =\case12 (\eta+\eta_\|) 
\end{equation} 
holds. Furthermore, let $\gamma_{11}$, $\gamma_1$, $\beta_s$ and 
$\gamma_s$ denote the critical indices that characterize, at $\rho=c=0$, 
the thermal singularities of the surface susceptibilities 
$\chi_{11}\equiv\chi_{11}(\bm{0})$ and $\chi_1\equiv\chi_1(\bm{0})$, 
the surface excess order parameter 
\begin{equation} 
  \label{eq:ms} 
  m_s=\int_0^\infty [G^{(0,1)}(\bm{r},z)-G^{(0,1)}(\bm{r},\infty)] 
  \,\rmd z 
\end{equation} 
and the surface excess susceptibility $\chi_s\equiv \rmd m_s/\rmd 
h|_{h=0}$, its derivative with respect to a bulk magnetic field $h$, 
via $\chi_{11}^{\rm sing}\sim g_\tau^{-\gamma_{11}}$, $\chi_1^{\rm 
  sing} \sim g_\tau^{-\gamma_1}$, $m_s\sim g_\tau^{\beta_s}$ and 
$\chi_s^{\rm sing} \sim g_\tau^{-\gamma_s}$, respectively. From 
equations (\ref{eq:GNMscf}) and (\ref{eq:G11decay}) we see that the 
scaling laws 
\begin{equation} 
  \label{eq:gamma11} 
  \gamma_{11}=\nu\,(1-\eta_\|)\;,\qquad 
  \gamma_1=\nu\,(2-\eta_\perp)\;,\\ 
\end{equation} 
and 
\begin{equation} 
\label{eq:betas} 
\beta_s=\beta-\nu\;,\qquad 
\gamma_s=\gamma+\nu\;,  
\end{equation} 
also remain valid for $m>0$.  
 
Finally, let us briefly consider the surface energy density 
$\langle[\phi^{\mathfrak B}(\bm{r})]^2\rangle$. In analogy to the CP 
case  
(Dietrich and Diehl 1981), the renormalization of this quantity involves 
further (additive) counter-terms, and the implied RG equation becomes 
inhomogeneous. We refrain from working out the details here, noting 
that the form 
\begin{equation} 
  \label{eq:sed} 
  \langle[\phi^{\mathfrak B}(\bm{r})]^2\rangle^{\rm sing}\sim 
  g_\tau^{2-\alpha-\nu-\Phi}\qquad(g_c=\rho=0) 
\end{equation} 
of its thermal singularity at $\rho=c=0$ can be inferred from equation 
(\ref{eq:GNMscf}). 
 
Next, we turn to the $\epsilon$-expansion of the surface critical 
exponents of the special transition. From equation (\ref{eq:eta1res}) 
we see that the $\Or(\epsilon)$ contribution to the fixed-point value 
$\eta_1^*$ vanishes for all $m>0$, so that 
\begin{equation}  
  \label{eq:eta1star}  
  \eta_\|^{\rm sp}-\eta=\eta_1^*=\Or(\epsilon^2) \qquad (m>0)\;. 
\end{equation} 
On the other hand, substitution of the fixed-point values 
(\ref{eq:ustar}) and (\ref{eq:lambdaplusstar}) into equation 
(\ref{eq:etacres}) yields 
\begin{equation}  
  \label{eq:Phiovnu}  
  \frac{\Phi}{\nu}-1=\eta_c^*= -\frac{n+2}{n+8}\,  
  \big[1+\lambda_0(m)\, 
  i'_1(\lambda_0;m)\big]\epsilon+\Or(\epsilon^2)\;. 
\end{equation} 
 
Let us see what happens to these results in the limit $m\to 0$.
Equation~(\ref{eq:Phiovnu}) reduces indeed for $m\to 0$ to the correct
$\Or(\epsilon)$ result for the CP case, just as the
$\epsilon$-expansions of the bulk critical exponents
(Diehl and Shpot 2000, Shpot and Diehl 2001) and the 
surface critical exponents of the LP ordinary transitions 
(Diehl \etal 2003a, 2003b) do. On the other hand, the 
$\Or(\epsilon)$ term of the exponent $\eta_\|^{\rm sp}$ of the CP 
special transition is known to be \emph{nonzero} 
(Bray and Moore 1977, Diehl and Dietrich 1981, 
1983). Consequently the limits $m\to 0$ of the $\epsilon$-expansion 
(\ref{eq:eta1star}) of $\eta_\|^{\rm sp}$ (as well as those of 
related surface exponents such as $\beta_1^{\rm sp}$) \emph{differ} 
from those of their $m=0$ counterparts for the CP special transition. 
Nevertheless, this is no cause for concern. The obvious reason for the 
apparent discrepancy between the $m\to 0$~limit of our result for 
$\eta_\|^{\rm sp}$ and the established one for $m=0$ is that the 
limits $\lambda\to \lambda_+^*(m)$ and $m\to 0$ do not commute. 
Taking the limit $m\to 0 $ last means to let $m\to 0$ while $\lambda$ 
follows the root $\lambda_+^*(m)$ on its way to infinity.  Since 
$\eta_1[u^*,\lambda_+^*(m)]$ is of order $\epsilon^2$, so is its limit 
${m\to 0}$. 
 
However, in order to compare with the CP case, the relevant problem to 
consider is the following. Given a generic initial value 
$\bar{\lambda}(1)\equiv \lambda<\infty$, what asymptotic behaviour 
does the RG flow for $m\to 0$ yield on large length scales? Noting 
that the running variable $\bar{\lambda}(\ell)\to \infty$, we see that 
the appropriate limit of $\eta_1$ that matters is 
$\lim_{\bar{\lambda}\to\infty}\lim_{m\to 0}\eta_1(u^*,\bar{\lambda})$. 
This implies the replacement of the integral $i_1(\lambda;m)$ by its 
${m\to 0}$~limit $i_1\equiv 1$ in our result (\ref{eq:eta1res}) for 
$\eta_1(u,\lambda)$, whereby it reduces to the correct one-loop 
expression for the CP case. Upon insertion of its value at $u=u^*$ 
into equation (\ref{eq:eta1star}), the familiar result for the 
exponent $\eta_\|^{\rm sp}$ of the CP special transition is recovered 
to order $\epsilon$. Thus, extrapolated to $m=0$, the RG flows 
obtained here yield results in conformity with the CP case, despite 
the mentioned difference between the ${m\to 0}$~limit of 
equation~(\ref{eq:eta1star}) and the $\epsilon$-expansion of 
$\eta_\|^{\rm sp}$ for the CP case. 
 
Let us try to use the results (\ref{eq:eta1star}) and 
(\ref{eq:Phiovnu}) to estimate the values of the associated surface 
critical exponents of the three-dimensional ANNNI model. Since we know 
these series expansions merely to first order in $\epsilon$, only 
crude estimates are possible. Clearly, it is desirable to exploit as 
much as possible the improved knowledge about the bulk critical 
exponents gained in  
Diehl and Shpot (2000), Shpot and Diehl (2001). It appears 
preferable to focus directly on estimates of the differences 
$\eta_\|^{\rm sp}-\eta$ or $\beta_1^{\rm sp}-\beta$ (both of which are 
proportional to $\eta_1^*$) and the ratio $\Phi/\nu $, rather than 
extrapolating the expansions to $\Or(\epsilon)$ of surface critical 
indices such as $\eta_\|^{\rm sp}$, $\beta_1^{\rm sp}$ or surface 
susceptibility exponents and $\Phi$. 
 
From the result (\ref{eq:eta1star}) we conclude that the differences 
${\eta_\|^{\rm sp}-\eta}$ and ${\beta_1^{\rm sp} -\beta}$ are likely 
to be small.  Accepting the best estimate $\eta\simeq 0.124$ of 
Shpot and Diehl (2001) for the uniaxial scalar case in $d=3$ 
dimensions, we conclude that $\eta_\|^{\rm sp}\simeq 0.1$. One way of 
estimating $\beta_1^{\rm sp}$ is to employ the values $\nu\simeq 
0.746$, $\theta=\nu_{l4}/ \nu \simeq {0.348}/{0.746}\simeq 0.47$ of 
Shpot and Diehl (2001) in equation (\ref{eq:beta1}), along with the 
approximation $\eta_\|^{\rm sp}\simeq \eta\simeq 0.126$. This gives 
$\beta_1^{\rm sp}\simeq 0.22$. If the slightly bigger value 
$\theta\simeq 0.487$ quoted in  
Diehl (2002) (and found by direct 
extrapolation of the $\epsilon$-expansion of $\theta$) is utilized 
instead, one obtains $\beta_1^{\rm sp}\simeq 0.23$. The agreement with 
Pleimling's Monte Carlo estimate $\beta_1^{\rm sp}=0.23(1)$ is 
impressive.  Another possibility is to start from the second part of 
equation (\ref{eq:beta1}) and utilize the estimate $\beta_1^{\rm sp} 
\simeq\beta$. The best estimate of  
Shpot and Diehl (2001) for $\beta$ was $\beta\simeq 0.246$. The most 
recent Monte Carlo result $\beta=0.238\pm 0.005$ 
(Pleimling 2002,  2004, Pleimling and Henkel 2001) is in 
conformity with this and lends support to the approximation 
$\beta\simeq \beta_1^{\rm sp}$. 
 
Unfortunately, we are not aware of any estimates of the surface 
crossover exponent $\Phi$ at $d=3$.  If we take again $m=n=1$ and 
boldly set $\epsilon=3/2$ in the expansion (\ref{eq:Phiovnu}), 
truncated at $\Or(\epsilon)$, we obtain $\Phi/\nu\simeq \case13 \times
1.65816\times \case32 \simeq 0.83$.  Utilizing once more the value 
$\nu\simeq 0.746$ then yields $\Phi\simeq 0.62$. We must caution the 
reader, however, to take these estimates \emph{cum grano salis}: the 
$\epsilon$-expansion to first order does not in general give reliable 
results.  Moreover, in the CP case (${m=0}$), the $\epsilon^2$ term of 
the crossover exponent $\Phi$ is known to have a rather large 
coefficient  
(Diehl and Dietrich 1981, 1983), a fact which makes 
it difficult to obtain precise estimates even from the 
$\epsilon$-expansion to second order.  Borel-Pad{\'e} estimates based 
on the massive field theory approach at fixed $d=3$  
(Diehl and Shpot 1994, 1998) give values 
$\Phi_{d=3,m=0,n=1}\simeq 0.54$ considerably smaller than the original 
one ($\simeq 0.68$) found in  
Diehl and Dietrich (1991, 1983), Diehl (1986) by 
evaluating the $\epsilon$-expansion to second order at $d=3$, though 
in much better agreement with Monte Carlo results 
(Landau and Binder 1980, Ruge \etal 1992, 1993). 
 
In extrapolating our $\epsilon$-expansion results for the scalar 
uniaxial case $m=n=1$ to three dimensions, we took it for granted that 
an LP special transition is possible at $d=3$. Clearly, a transition to 
a surface phase with long-range order should be possible at 
temperatures $T_s$ higher than the line of bulk critical temperatures 
$T_c$ if $m=n=1$. Since $T_s$ can be varied through a change of 
surface interaction constants, one expects that for appropriately 
fine-tuned surface enhancement $T_s$ can become equal to the LP 
temperature, so that a special LP transition ought to be possible at 
$d=3$, as also recent Monte Carlo results  
(Pleimling 2002) indicate. 
 
For other values of $m$ and $n$ extrapolations to $d=3$ make little 
sense. Since we precluded in the Hamiltonian 
(\ref{eq:Hamf})--(\ref{eq:L1}) any (bulk and surface) terms breaking 
its $O(n)$ symmetry, a bulk-disordered surface phase with long-range 
order---and hence a special transition of the kind considered 
above---cannot occur for $d=3$ if $n>1$ because it would require the 
breaking of a continuous symmetry in an effectively two-dimensional 
system. 
 
It is tempting to anticipate the possibility of a special transition 
for the biaxial scalar case $m=2$, $n=1$. However, in the case of a 
multi-axial LP, one generically expects contributions to the 
Hamiltonian that break its rotational invariance in the 
($m>1$)-dimensional subspace of $\alpha$-coordinates, such as bulk 
terms of cubic symmetry, $\sum_{\alpha=1}^n(\partial_\alpha^2\phi)^2$, 
and similar space anisotropies of further reduced symmetry. According 
to a recent two-loop RG analysis  
(Diehl \etal 2003c), such anisotropies are relevant, at least for 
small $\epsilon>0$. Unfortunately, no stable new (anisotropic) fixed 
point could be found.  To our knowledge, Monte Carlo investigations of 
appropriate biaxial generalization of the $d=3$ ANNNI model have not 
yet been carried out. Thus, whenever such anisotropies cannot be 
ruled out, it is presently unclear whether a biaxial bulk LP exists, 
in particular in three dimensions. The clarification of this issue is 
beyond the scope of the present paper.  Note also, that in order to 
account for the absence of rotational symmetry in the $m$-dimensional 
subspace, we would have to generalize the surface part ${\mathcal 
  L}_1$ of the Hamiltonian by allowing its derivative term to become 
anisotropic as well. 
 
\section{RG~improved perturbation theory} 
\label{sec:RGipt}  
 
In the previous section we exploited the RG equations (\ref{eq:RGE}) 
to deduce the scaling forms (\ref{eq:GNMscf}) of the ($N+M$)-point 
cumulants (\ref{eq:GNM}). A tacit assumption underlying this 
derivation is that the renormalized cumulants $G^{(N,M)}_{\rm ren}$ 
are well-behaved at the fixed point ${\mathcal P}^*_{\rm sp}$. More 
precisely, these functions were assumed to approach for $\epsilon>0$ 
finite and nonzero limits, as $\lambda$ and $u$ approach their 
respective fixed-point values $\lambda_+^*$ and $u^*$. This can and 
should be checked within the framework of RG~improved perturbation 
theory. Furthermore, other conditions exist to which this statement 
applies just as much. For example, consistency of the scaling forms 
for $g_\tau \neq 0$ with their counterparts for $g_\tau =0$ imposes 
constraints on the asymptotic $g_\tau$-dependence of the scaling 
function for $g_\tau\to 0$.  Further constraints concern the 
dependence on $g_c$; they arise from matching requirements between the 
$g_c$-dependent scaling forms and the asymptotic behaviour at those 
transitions to which the special transition crosses over, namely the 
ordinary, surface or extraordinary transition. 
 
Our aim here is the explicit verification of some of these properties 
by means of RG~improved perturbation theory to one-loop order. For the 
sake of simplicity, we will restrict ourselves to the illustrative 
case of the (renormalized) surface susceptibility $\chi^{(\rm 
  ren)}_{11}(\bm{p})$. We begin by considering this quantity at 
${\tau=\rho =0}$ for general nonzero momentum $\bm{p}=(p_\alpha 
,p_\beta )\in\mathbb{R}^{d-1}$ as function of $g_c$. Upon introducing 
the dimensionless momenta 
\begin{equation} 
  \label{eq:pPdef} 
  \hat{p}=\big(p_\beta  p_\beta  / \mu^2\big)^{1/2}\;,\qquad  
\hat{P}={\big(\sigma^{1/2} p_\alpha p_\alpha/ \mu\big)}^{1/2}\;, 
\end{equation} 
we can write its scaling form as 
\begin{eqnarray} 
  \label{eq:chi11scf} 
  \chi^{(\rm ren)}_{11}(\bm{p})&\approx& \mu^{-1}\, 
  \hat{p}^{\eta_\|^{\rm sp}-1}\,  
  \Xi{\big(\hat{P} \hat{p}^{-\theta },g_c\, \hat{p}^{-\Phi/ \nu 
      }\big)} \\  
&=&\mu^{-1}\,{\hat{P}}^{\big(\eta_\|^{\rm sp}-1\big)/ \theta}\, 
\Psi{\big(\hat{P}  
    \hat{p}^{-\theta },g_c\,\hat{p}^{-\Phi/ \nu }\big)}\;,  
\end{eqnarray} 
where the scaling functions $\Psi$ and $\Xi$ are related to each other 
via 
\begin{equation} 
  \label{eq:Xixi1} 
  \Psi({\mathsf P},{\mathsf c})= {\mathsf P}^{\big(1-\eta_\|^{\rm 
      sp}\big)/\theta}\,  
  \Xi({\mathsf P},{\mathsf c}) \;. 
\end{equation} 
 
Consistency with the scaling forms that hold if either $\hat{p}=0$ or 
$\hat{P}=0$ requires the limiting behaviour 
\begin{equation} 
  \label{eq:Xias} 
  \Xi({\mathsf P},{\mathsf c})\approx \cases{\Xi_0({\mathsf c})& 
    for ${\mathsf P}\to 0$,\cr  
  {\mathsf P}^{\big(\eta_\|^{\rm sp}-1\big)/ \theta}\, 
    \Psi_\infty({\mathsf c})&for ${\mathsf P}\to \infty$.} 
\end{equation} 
Next we consider the limits $\hat{p}\to 0$ and $\hat{P}\to 0$ for 
$g_c>0$. The limiting dependences on $\hat{p}$ and $\hat{P}$ must be 
in conformity with the leading infrared singularities $\sim 
\hat{p}^{\eta_\|^{\rm ord}-1}$ and $\sim \hat{P}^{(\eta_\|^{\rm 
    ord}-1)/\theta}$ at the ordinary transition.  Hence we anticipate 
that $\Xi$ varies for ${\mathsf c}\to \infty$ as 
\begin{equation} 
  \label{eq:Xiasord} \fl
  \Xi({\mathsf P},{\mathsf c}) \mathop{\approx}_{{\mathsf c}\to\infty} 
{\mathsf c}^{-\nu\big(1-\eta_\|^{\rm 
      sp}\big)/\Phi}\big[X_{11}+Y_1\,{\mathsf P}^2\,{\mathsf 
    c}^{-2\nu\theta/\Phi}+\ldots\big] + 
  {\mathsf c}^{-\nu\big(\eta_\|^{\rm ord}-\eta_\|^{\rm 
      sp}\big)/\Phi}\,\Xi_\infty({\mathsf P}) 
+\ldots, 
\end{equation} 
where $X_{11}$ and $Y_1$ are constants, while $\Xi_\infty({\mathsf P})$ 
behaves as 
\begin{equation} 
  \label{eq:Xiinftyas} 
  \Xi_\infty({\mathsf P}) \approx \cases{\Xi_{\infty,0}\equiv 
    \Xi_\infty(0)&for ${\mathsf P}\to 0$,\cr  
\Xi_{\infty,\infty}\,{\mathsf P}^{\big(\eta_\|^{\rm ord}-1\big)/\theta}& 
  for ${\mathsf P} \to \infty $.}  
\end{equation} 
The latter properties ensure that the term of
equation~(\ref{eq:Xiasord}) following the square brackets produces the
required powers $\hat{p}^{\eta_\|^{\rm ord}-1}$ and
$\hat{P}^{(\eta_\|^{\rm ord}-1)/\theta}$.  To understand the terms of
equation~(\ref{eq:Xiasord}) involving the square brackets, one should
recall that $\chi_{11}$ remains \emph{finite} at the ordinary
transition.  Thus a momentum-\emph{independent} contribution to
$\chi_{11}^{\rm (ren)}(\bm{0})$ must exist, apart from additional ones
with analytic momentum dependence. The contribution ${\propto X_{11}}$
accounts for the former; except for a contribution proportional to
$\hat{P}^2$, terms of the latter analytic kind have been suppressed,
along with momentum-dependent subleading corrections.
 
In \ref{app:surfsusc} the renormalized function $\chi_{11}^{(\rm ren)}$ 
is computed to one-loop order for $\tau=\rho=0$ and general values of 
$c\geq0$ and $\lambda\geq 0$. The result can be written as 
\begin{eqnarray} 
  \label{eq:chi11pt}\fl 
  \frac{\mu}{\chi_{11}^{(\rm ren)}(\bm{p})} 
  =\kappa_{\bm{p}}+c+\lambda\,\hat{P}^2 
+\frac{u}{2}\,\frac{n+2}{3}\, 
  \bigg\{\big[(\kappa_{\bm{p}}+c\big)\, i_1 
 +c\,\big(1+\lambda\,i_1')\big] 
 \ln(2\kappa_{\bm{p}})-c\lambda\,i_1^{(1,1)} 
\nonumber\\ \lo\strut  
-(\kappa_{\bm{p}}+c)\,i_1^{(0,1)} 
+2\,\frac{c^2}{\hat{p}}\,{\mathcal  
  A}{\big(\hat{P}/\sqrt{\hat{p}},c/\hat{p},\lambda\big)} 
+\Or(\epsilon)\bigg\}+\Or(u^2)\;,  
\end{eqnarray} 
where 
\begin{equation} 
  \label{eq:kappa} 
  \kappa_{\bm{p}}\equiv\sqrt{\hat{p}^2+{\hat{P}}^4}\;, 
\end{equation} 
while the function ${\mathcal A}$ is an average of the form 
(\ref{eq:aveps}): 
\begin{eqnarray} 
  \label{eq:Av} 
  {\mathcal A}({\mathsf P},{\mathsf c},\lambda)\equiv  
\left\langle\frac{\ln\big[{\mathsf 
      c}/(1+\lambda t)\sqrt{1+{\mathsf P}^4}\,\big]}{(1+\lambda 
    t)^2\big[{\mathsf 
      c}-(1+\lambda t)\sqrt{1+{\mathsf P}^4}\,\big]} \right\rangle_{0,m}\;. 
\end{eqnarray} 
Further, $i_1^{(0,1)}$ and $i_1^{(1,1)}$ denote the partial 
derivatives $\partial i_1/\partial \epsilon$ and 
$\partial^2i_1/\partial \epsilon\partial\lambda$, respectively. Just 
as $i_1$ and $i_1'$, these are to be evaluated at $\epsilon=0$ in 
equation~(\ref{eq:chi11pt}). 
 
From the perturbative result (\ref{eq:chi11pt}), the universal scaling 
function $\Xi({\mathsf P},{\mathsf c})$ must follow up to non-universal 
metric factors upon setting the coupling constants $u$ and $\lambda$ 
to their fixed-point values (\ref{eq:ustar}) and 
(\ref{eq:lambdaplusstar})%
\footnote{As non-universal factors we have, 
  in this case, the overall amplitude of $\Xi({\mathsf P},{\mathsf 
    c})$ and the two metric factors associated with $\sigma$ and 
  $g_c=c$, respectively.}. 
Recalling the $\Or(\epsilon)$ expressions (\ref{eq:eta1star}) and 
(\ref{eq:Phiovnu}) for $\eta_\|^{\rm sp}$ and $\Phi/\nu$, one easily 
verifies that $\chi_{11}^{\rm (ren)}$ takes indeed the scaling 
form~(\ref{eq:chi11scf}), where the scaling function is given by 
\begin{eqnarray} 
  \label{eq:Xi} 
\fl 
  \frac{1}{\Xi({\mathsf P},{\mathsf c})}=a_n(\epsilon)\, 
  \sqrt{1+{\mathsf P}^4} +b_n(\epsilon)\,{\mathsf c}  
  +\lambda_0\,{\mathsf P}^2 
 +\frac{n+2}{n+8}\,\epsilon\Big\{2\, {\mathsf c}^2\,{\mathcal 
   A}({\mathsf P},{\mathsf c},\lambda_0) 
\nonumber\\ \strut 
+{\mathsf c}\, \big[1+\lambda_0\,{i_1'(\lambda_0;m)}\big] \,\ln 
\Big[2\sqrt{1+{\mathsf P}^4}\Big]\Big\}+\Or(\epsilon^2)  
\end{eqnarray} 
with 
\begin{equation} 
  \label{eq:an} 
  a_n(\epsilon)=1-\epsilon\,\frac{n+2}{n+8}\, 
  i_1^{(0,1)}(\lambda_0,0;m) 
\end{equation} 
and 
\begin{equation} 
  \label{eq:bn} 
b_n(\epsilon)=1-\epsilon\,\frac{n+2}{n+8}\, 
\Big[i_1^{(0,1)}(\lambda_0,0;m) + 
\lambda_0\,i_1^{(1,1)}(\lambda_0,0;m)\Big]\;. 
\end{equation} 
 
Obviously the limiting forms (\ref{eq:Xias}) for small and large 
${\mathsf P}$ hold. Checking the consistency with the critical 
behaviour at the ordinary transition is less trivial because the 
$\epsilon$-expanded version (\ref{eq:Xi}) of the scaling function does 
not yet contain the anticipated asymptotic power laws (\ref{eq:Xias}) 
and (\ref{eq:Xiasord}) in exponentiated form. To proceed, we consider 
the behaviour of $\Xi^{-1}$ as ${\mathsf c}\to \infty$. The leading 
contributions are of order ${\mathsf c}\ln{\mathsf c}$. One finds that 
the terms $\propto \ln\sqrt{1+{\mathsf P}^4}$ of this order cancel. 
The remaining (momentum-independent) terms $\sim {\mathsf 
  c}\ln{\mathsf c}$ can be combined with part of the contributions 
$\sim {\mathsf c}$ to identify the power ${\mathsf c}^{1-\eta_c^*}$ to 
$\Or(\epsilon)$. The associated exponent satisfies $1-\eta_c^*=\nu 
(1-\eta_\|^{\rm sp})/\Phi +\Or(\epsilon^2)$. Hence this contribution 
gives rise to the first term in equation~(\ref{eq:Xiasord}). For its 
amplitude $X_{11}$ we obtain the $\epsilon$-expansion 
\begin{eqnarray}\fl 
  \label{eq:X11} 
  X_{11}^{-1}=1 -\epsilon\,\frac{n+2}{n+8}\, 
  \bigg\{i_1^{(0,1)}(\lambda_0,0;m)+\lambda_0\,i_1^{(1,1)}(\lambda_0,0;m) + 
  2\left\langle\frac{\ln(1+\lambda_0
       t)}{(1+\lambda_0 t)^2}\right\rangle_{0,m} 
\nonumber\\ \lo \strut-\big[1+\lambda_0\,i_1'(\lambda_0;m)\big]\ln 2\bigg\} +\Or(\epsilon^2)\;. 
\end{eqnarray} 
  
Turning to the contributions of order $\ln{\mathsf c}$ and ${\mathsf 
  c}^0$, we note that the former can be combined with part of the 
latter to obtain the term $\epsilon\case{n+2}{n+8}\,\kappa 
\ln({\mathsf c}/\kappa)$, where $\kappa\equiv \sqrt{1+{\mathsf P}^4}$, 
and we employed the fact that $\langle(1+\lambda_0)^{-1}\rangle_{0,m}= 
(i_1(\lambda_0;m)+\langle1\rangle_{0,m})/2=1/2$. This logarithm can be 
cast in the form $\kappa\,[({\mathsf 
  c}/\kappa)^{\epsilon\,(n+2)/(n+8)}-1]+\Or(\epsilon^2)$. Recalling the 
$\epsilon$-expansion 
\begin{equation} 
  \label{eq:etaparord} 
  \eta_\|^{\rm ord}=2-\epsilon\,\frac{n+2}{n+8}+\Or(\epsilon^2)\;, 
\end{equation} 
one easily verifies that the term $\kappa \,({\mathsf 
  c}/\kappa)^{\epsilon\,(n+2)/(n+8)}$ produces a contribution to $\Xi$ 
consistent with the form of the  term in the second line of 
equation~(\ref{eq:Xiasord}), with 
\begin{eqnarray} 
  \label{eq:Xiinf} 
  \Xi_\infty({\mathsf P})=-\left(1+{\mathsf P}^4\right)^{(\eta_\|^{\rm 
      ord}-1)/4\theta}[1+\Or(\epsilon)]\;. 
\end{eqnarray} 
Furthermore, the term $\propto \lambda_0\,{\mathsf P}^2$ in 
equation~(\ref{eq:Xi}) yields an analytic contribution $\propto{\mathsf 
  P}^2$. Its form matches the one $\propto Y_1$ in 
equation~(\ref{eq:Xiasord}), where the  coefficient is  given by 
\begin{equation} 
  \label{eq:Y1} 
  Y_1=-\lambda_0+\Or(\epsilon)\;. 
\end{equation} 
 
In summary, we find that our result (\ref{eq:Xi}) for the scaling 
function $\Xi$ is also in conformity with the limiting forms 
(\ref{eq:Xiasord}) and (\ref{eq:Xiinftyas}). This means that the 
asymptotic momentum singularities one obtains at the ordinary 
transition from the $c$-dependent analysis here, to first order in 
$\epsilon$, agree with those found in  
Diehl \etal (2003a) for the case $\mathring{c}=\infty$ with Dirichlet 
boundary conditions. While expected, this manifestation of 
universality is gratifying in that it corroborates our identification 
of the fixed points ${\mathcal P}^*_{\rm sp}$ and ${\mathcal P}^*_{\rm 
  ord}$. 
 
\section{Summary and conclusions}  
\label{sec:concl} 
 
We have investigated the special surface transition that occurs at an 
$m$-axial bulk LP in the case where the $m$ (potential) modulation 
axes are parallel to the surface. Our RG analysis in 
$d=4+\case{m}{2}-\epsilon$ dimensions corroborates the observation of 
Diehl \etal (2003a, 2003b) that a square gradient term of 
the form shown in equation (\ref{eq:L1}) must be included in the 
boundary part ${\mathcal L}_1$ of the Hamiltonian, in addition to its 
usual $\phi^2$ contribution. 

The RG fixed point associated with the special transition is located 
at a nontrivial value $\lambda_+^*=\Or(\epsilon^0)$ of the associated 
renormalized surface coupling constant $\lambda$, zero (renormalized) 
surface enhancement $c=0$ (see equation (\ref{eq:Pfix})) and the usual 
fixed-point value $u^*$ of order $\epsilon$ of the bulk interaction 
constant. As a direct consequence of the fact that 
$\lambda_+^*=\Or(\epsilon^0)$, the surface crossover exponent $\Phi$ 
becomes $m$~dependent already at \emph{first} order in $\epsilon$, 
whereas this happens for the bulk critical indices and the surface 
critical exponents of the ordinary transition only at \emph{second} 
order in $\epsilon$. 
 
It would be natural to anticipate the same behaviour as that of $\Phi$
for the surface correlation exponent $\eta_\|^{\rm sp}$, the second
independent surface critical index of the special transition. Yet this
is \emph{not} the case because the expansion to order $\epsilon$ of
$\eta_\|^{\rm sp}$ turns out to \emph{vanish}, whereas the
contribution of order $\epsilon$ of its $m=0$ analogue is known to be
nonzero
(Diehl and Dietrich 1981, 1983), as is its $\Or(\epsilon^2)$ term.
Thus the limit $m\to 0$ of the $\epsilon$-expansion of $\eta_\|^{\rm
  sp}$ differs from the established $m=0$ result for the CP case. This
behaviour is exceptional. For once, the $\epsilon$-expansions of all
those \emph{bulk} critical indices that retain their physical meaning
for $m=0$ turn over into their $m=0$~counterparts for the CP case in
the limit $m\to 0$, as has been explicitly verified to order
$\epsilon^2$
(Shpot and Diehl 2001). Second, the same applies to the surface 
critical exponents of the \emph{ordinary} transition 
(Diehl \etal 2003a, 2003b).  Moreover, even our 
$\Or(\epsilon)$ result for $\Phi$ implied by equation 
(\ref{eq:Phiovnu}) is fully consistent with the $\epsilon$-expansion 
of $\Phi_{m=0}$  
(Diehl and Dietrich 1981, 1983). 
 
As we have seen, the origin of the discrepancy between the 
$\epsilon$-expansions of $\lim_{m\to 0}\eta_\|^{\rm sp}$ and its $m=0$ 
analogue for the CP case can be traced back to the non-commutativity 
of the limits $\lambda\to\lambda_+^*(m)$ and $m\to 0$ of the 
RG~function $\eta_1(u^*,\lambda)$.  It is gratifying that we could 
recover the correct $\Or(\epsilon)$ term of $\eta_\|^{\rm sp}$ for the 
CP case by analysing the RG flow directly for $m\to 0$. 
 
The above, not necessarily expected, results are a clear message that 
further surprises might well be encountered when the surface 
orientation is taken to be perpendicular. Since in this case the 
distance $z$ from the surface scales naively as $\mu^{-1/2}$ (rather 
than as the inverse of the momentum scale $\mu$), more surface 
monomials exist whose coupling constants have nonnegative momentum 
dimensions for $\epsilon \geq 0$, and hence are potentially dangerous. 
This makes the analysis of multi-critical surface transitions such as 
the special one even more interesting and challenging. 
 
The series expansions of the surface critical exponents of the LP 
special transition determined here are restricted to first order in 
$\epsilon$. Unless their knowledge can be combined with information 
from other sources, one certainly cannot hope to get numerically 
accurate estimates of the surface critical exponents at ${d=3}$. The 
additional information we could benefit from was the 
$\epsilon$-expansions of the bulk critical exponents to 
$\Or(\epsilon^2)$ and previous field-theory and Monte Carlo estimates 
for $d=3$  
(Shpot and Diehl 2001, Pleimling and Henkel 2001). Our result 
(\ref{eq:eta1star}), which means that $\beta_1^{\rm 
  sp}=\beta+\Or(\epsilon^2)$, is in accordance with the small 
difference between the values of $\beta_1^{\rm sp}$ and the bulk 
exponent $\beta$ obtained in  
Pleimling (2002) via Monte Carlo simulations of the three-dimensional 
ANNNI model. 
 
We see no principal obstacle to extending the present analysis to 
second order in $\epsilon$. However, the effort required appears to be 
considerably greater than was necessary for our---already quite 
demanding---analysis of the ordinary transition  
(Diehl \etal 2003a, 2003b). 
Furthermore, the residues of some pole terms one must determine seem to 
be expressible only in terms of multi-dimensional integrals which must 
be computed by numerical means. 
 
In order to corroborate our identification of the fixed points 
(\ref{eq:Pfix}) we also computed the scaling functions (\ref{eq:Xi}) 
of the momentum-dependent local surface susceptibility 
$\chi_{11}(\bm{p})$ at the Lifshitz point ${\tau=\rho=0}$, to first order in 
$\epsilon$, and confirmed that the momentum singularities at both the 
special as well as the ordinary transition have the predicted 
power-law forms, consistent with our own RG results here and those of 
Diehl \etal (2003a) for the ordinary transition. 
 
We close with three remarks.  
\begin{list}{(\roman{enumi})}{\usecounter{enumi}}
\item There are several worthwhile goals which future Monte Carlo
  simulations of the ANNNI model might accomplish.  An obvious one is
  obtaining an accurate estimate of the surface crossover exponent
  $\Phi$. A second---equally important and not unrelated---task is a
  careful check of the predicted scaling behaviour at the special
  transition. We believe that appropriate (non)linear scaling fields
  $g_c$, $g_\tau$ etc [see equations (\ref{eq:gtaucintro})] should be
  introduced when performing the required analysis of the data.
\item While mathematically rigorous results on the surface phase
  diagram of the three-dimensional semi-infinite Ising model can be
  found in the literature
  (Fr{\"o}hlich and Pfister 1987, Pfister and Penrose 1988), we are
  not aware of similar work on the semi-infinite ANNNI model. Clearly,
  rigorous results both on bulk and surface phase diagrams involving
  bulk LP could be very valuable.
\item Finally, let us emphasize the need for careful experimental
  studies of bulk and surface critical behaviour at {LP}. Owing to the
  enormous advances in experimental technology and our better
  theoretical understanding, tests much more stringent than the (more
  than 20-year-old) experimental study
  (Shapira \etal 1981) of bulk critical behaviour at the uniaxial LP
  of MnP ought to be possible today. The situation is even worse for
  \emph{surface} critical behaviour at LP, since experimental
  investigations do not yet exist apparently.
\end{list}
 
\ackn 
 
We gratefully acknowledge partial support by the Deutsche 
Forschungsgemeinschaft (DFG) via grant Di-378/3. 
 
\appendix 
 
\section{Calculation of $A_c$}\label{app:Ac} 
   
In this appendix we derive the one-loop result~(\ref{eq:Ac}) for the 
renormalization function $A_c$. To this end, we consider the surface 
cumulant $G^{(0,2)}_{\rm ren}(\bm{p})$ for ${\tau=c=0}$ and generic 
$\lambda >0$ and $\rho >0$. The free propagator of the renormalized 
theory may be gleaned from equation~(7) of  
Diehl \etal (2003a) and the 
re-parametrizations (\ref{eq:bulkrep}) and (\ref{eq:surfrep}).  We 
express it in terms of the dimensionless momenta $\hat{p}$ and 
$\hat{P}$ introduced in equation (\ref{eq:pPdef}). For $\tau=c=0$, it 
then becomes 
\begin{equation} 
  \label{eq:freeG} 
\hat{G}(\bm{p};z_1,z_2) = 
\frac{1}{2\mu \kappa_{\bm{p}}} 
  \bigg[\e^{-\kappa_{\bm{p}}|z_1-z_2|} 
+ \frac{\kappa_{\bm{p}}-\lambda\hat{P}^2}{\kappa_{\bm{p}}+ 
  \lambda\hat{P}^2}\,   
\e^{-\mu \kappa_{\bm{p}}(z_1+z_2)}\bigg] 
\end{equation} 
in the $\bm{p}z$~representation, with 
\begin{equation} 
\label{eq:kapparho} 
\kappa_{\bm{p}}=\sqrt{\hat{p}^2+\hat{P}^4+\rho \hat{P}^2}\;. 
\end{equation} 
Upon taking into account all counter-terms that contribute to first 
order in $u$, we obtain 
\begin{eqnarray} 
 \label{eq:G02lam}\fl 
\hat{G}_{\rm ren}^{(0,2)}(\bm{p})= 
\frac{Z_1^{-1}/\mu}{\kappa_{\bm{p}}+(\lambda+P_\lambda)\hat{P}^2}\nonumber\\ \lo 
 +{\int^\infty_{0-}}\rmd z \, 
 [\hat{G}(\bm{p};0,z)]^2 
 \bigg[\; 
\unitlength1mm 
\raisebox{-0.15em}{\begin{picture}(6,6) 
\thicklines 
\put(0,0){\line(1,0){5}} 
\put(2.5,2.5){\circle{5}} 
\put(1.7,-2.5){\small $\bm{x}$} 
\end{picture}} 
-A_\tau\mu^2\rho^2-A_c\mu\rho\,\delta(z) 
\bigg] 
+\Or(u^2)\,,\nonumber\\ 
\end{eqnarray} 
where the lower limit $0-$ of the $z$~integration serves to ensure 
that $\int\rmd z\,\delta(z)=1$.

The tadpole graph displayed in this equation yields two contributions: 
one corresponding to the translation invariant part of $G$---its 
`bulk' part $G_{\rm b}$---and a second one resulting from its 
remaining `surface' part $G_{\rm s}$. Since 
\begin{equation} 
  \label{eq:Gb} 
  G_{\rm b}(\bm{x}-\bm{x})= {\int}\frac{\rmd^{d-1}p}{(2 \pi)^{d-1}}\,  
 \frac{1}{2\mu\kappa}_{\bm{p}} = 
\frac{\mu^{2-\epsilon}\,F_{m,0}}{\sigma^{m/4}}\, 
\frac{m\rho^2}{8\epsilon} +\Or(\epsilon^0)\;, 
\end{equation} 
the pole term of \; 
\unitlength0.8mm 
\raisebox{-0.15em}{\begin{picture}(6,6) 
\thicklines 
\put(0,0){\line(1,0){5}} 
\put(2.5,2.5){\circle{5}} 
\put(1.7,5.4){\scriptsize b} 
\end{picture}} 
(the former) gets cancelled by the subtraction $\propto A_\tau$ in the 
square brackets, for our choice of $A_\tau$, made in accordance with 
equation~(17) of  
Diehl \etal (2003a). 
 
The graph 
\; 
\unitlength0.8mm 
\raisebox{-0.15em}{\begin{picture}(6,6) 
\thicklines 
\put(0,0){\line(1,0){5}} 
\put(2.5,2.5){\circle{5}} 
\put(1.7,5.4){\scriptsize s} 
\end{picture}} 
yields pole terms of the form $\delta'(z)/ \epsilon$ and 
$\delta(z)/\epsilon$.  To see this, note that upon making the change 
of variables $\hat{p}\to \hat{p}/2\mu z$ and $\hat{P}\to 
\hat{P}/\sqrt{2\mu z}$, the surface part $G_{\rm s}(\bm{x},\bm{x})$ 
can be written as 
\begin{equation} 
  \label{eq:Gs} 
  G_{\rm s}(\bm{x},\bm{x}) 
= \sigma^{-m/4}(2 z)^{\epsilon-2}f_\epsilon(\mu z) 
\end{equation} 
with  
\begin{equation} 
f_\epsilon(\hat{z})=\int\frac{\rmd^{d-m-1}\hat{p}\, 
  \rmd^m\hat{P}}{(2\pi)^{d-1}}\; 
\frac{\kappa(\hat{p},\hat{P},\hat{z}) 
  -\lambda\hat{P}^2}{\kappa(\hat{p},\hat{P},\hat{z})  
  +\lambda\hat{P}^2}\, 
\frac{\e^{-\kappa(\hat{p},\hat{P},\hat{z})}}{2\,\kappa(\hat{p},\hat{P},\hat{z})} 
\;, 
\end{equation} 
where  
\begin{equation} 
  \label{eq:kappapz} 
  \kappa(\hat{p},\hat{P},\hat{z})=\sqrt{\hat{p}^2 
    +\hat{P}^4+2\hat{z}\rho\hat{P}^2}\;. 
\end{equation} 
The pole terms of the generalized function (\ref{eq:Gs}) originate 
from the contributions that behave as $z^{\epsilon-2}$ and 
$z^{\epsilon-1}$ near $z=0$. Utilizing the well-known Laurent 
expansion $z^{\epsilon-k}=(-1)^k\,\delta^{(k)}(z)/k!+\Or(\epsilon)$ of 
such generalized functions  
(Gelf'and and Shilov 1964) gives 
\begin{equation} 
  \label{eq:fexp} 
  (2\hat{z})^{\epsilon-2}\,f_\epsilon(\hat{z})= 
  -\frac{f_0(0)}{4\epsilon}\,\delta'(\hat{z})  
  +\frac{f'_0(0)}{4\epsilon}\,\delta(\hat{z}) +\Or(\epsilon)\;. 
\end{equation} 
The Taylor coefficients $f(0)$ and $f'(0)$ can both be expressed in 
terms of the function $i_1$ and its derivative $i_1'$. A 
straightforward calculation yields 
\begin{equation} 
\label{eq:f(0)} 
f_0(0)= 2 F_{m,0}\, i_1(\lambda;m) 
\end{equation} 
and 
\begin{equation}\fl 
\label{eq:fprime(0)}   
f_0'(0)=- 4\rho\,F_{m,0} \bigg[\frac{1-i_1 (\lambda;m) 
  -\lambda\,i'_1(\lambda;m)}{2\lambda}- \frac{\sqrt{\pi} 
  \,\Gamma[(m+2)/4)]}{2\,\Gamma({m}/{4})}\bigg]\;.    
\end{equation}  
The uv singularity of the graph  
\; 
\unitlength0.8mm 
\raisebox{-0.15em}{\begin{picture}(6,6) 
\thicklines 
\put(0,0){\line(1,0){5}} 
\put(2.5,2.5){\circle{5}} 
\put(1.7,5.4){\scriptsize s} 
\end{picture}} 
implied by the pole term $\propto \delta(z)$ in equation 
(\ref{eq:fexp}) is momentum independent; it must be cancelled by a 
renormalization of the surface enhancement $\mathring{c}$. Since we 
set the renormalized quantity $c$ to zero, this condition immediately 
gives the renormalization function $A_c$ to one-loop order. The result 
(\ref{eq:Ac}) can be read off from the above equations. The pole term 
$\propto\delta'(z)$ in equation (\ref{eq:fexp}) looks like a 
counter-term of the form 
$\int_{\mathfrak{B}}\bm{\phi}\partial_n\bm{\phi}$, which we did 
\emph{not} introduce, knowing that it can be transformed by means of 
the boundary condition 
\begin{equation} 
\label{eq:bc} 
\partial_n\bm{\phi}=(\mathring{c} 
-\mathring{\lambda}\,\partial_\alpha\partial_\alpha)\,\bm{\phi}  
\end{equation} 
into surface counter-terms we included, such as the one $\propto 
\int_{\mathfrak{B}}( \partial_\alpha\bm{\phi})^2$  
(Diehl \etal 2003a). In 
fact, using the one-loop results 
\begin{equation}\fl 
  \label{eq:Z1Plam} 
  Z_1(u,\lambda,\epsilon)-1=-\lambda^{-1}P_\lambda(u,\lambda,\epsilon) 
  +\Or(u^2)=\frac{n+2}{3}\,\frac{i_1(\lambda;m)\,u}{2\epsilon}+\Or(u^2) 
\end{equation} 
of this reference, one easily convinces oneself that the pole terms 
produced by the renormalization functions $Z_1$ and $P_\lambda$ of the 
first term in equation (\ref{eq:G02lam}) cancel those originating from 
the singularity $\propto \delta'(z)$ of $G_{\rm s}(\bm{x},\bm{x})$. 
 
\section{One-loop calculation of $\bm{\chi_{11}(\bm{p)}}$} 
\label{app:surfsusc} 
In this appendix we outline the computation of the renormalized 
surface susceptibility $\chi^{(\rm ren)}_{11}(\bm{p})$ to one-loop 
order for ${\tau=\rho=0}$ and general values $c\geq0$ and $\lambda>0$. 
The contribution of the bulk part  
\unitlength0.8mm 
\;\raisebox{-0.15em}{\begin{picture}(6,7) \thicklines 
    \put(0,0){\line(1,0){5}} \put(2.5,2.5){\circle{5}} 
    \put(1.7,5.4){\scriptsize b} 
\end{picture}}\; 
to the one-loop term vanishes for ${\tau=\rho=0}$ (in dimensional 
regularization), just as the terms proportional to $A_\tau$ and $A_c$ 
in equation (\ref{eq:G02lam}) do. In the remaining one-loop integral 
we perform both the integration over $z$ as well as the angular 
integrations, obtaining 
\begin{equation}\fl 
  \label{eq:chi11} 
 \frac{\mu}{\chi_{11}^{(\rm ren)}(\bm{p})}={Z_1}{\left[\kappa_{\bm{p}} 
    +(\lambda+P_\lambda)\hat{P}^2+Z_c\,c\right]} 
 +\frac{u}{2}\, \frac{n+2}{3}\,\frac{K_{d-m-1}K_m}{F_{m,\epsilon}}\,J 
 +\Or(u^2) 
\end{equation} 
with 
\begin{equation}\fl 
  \label{eq:Jdef} 
  J={\int_0^\infty}{\rmd}\hat{p}_1\, 
  \hat{p}_1^{d-m-2}{\int_0^\infty}{\rmd}\hat{P}_1\,\hat{P}_1^{m-1}  
\,\frac{\kappa_{\bm{p}_1}-c  
  -\lambda\hat{P}_1^2}{4\kappa_{\bm{p}_1}(\kappa_{\bm{p}} 
  +\kappa_{\bm{p}_1})(\kappa_{\bm{p}_1} +c+\lambda{\hat{P}_1}^2)}\;. 
\end{equation} 
Here $\hat{p}$ and $\hat{P}$ denote again the dimensionless momenta 
(\ref{eq:pPdef}), while $\kappa_{\bm{p}}$, defined by equation 
(\ref{eq:kappa}), is the $\rho=0$ analogue of the quantity 
(\ref{eq:kapparho}). The renormalization factor $Z_c$ may be gleaned 
from equation~(22) of  
(Diehl \etal 2003a): expressed in terms of $i_1'$, it 
reads 
\begin{equation} 
  \label{eq:Zc} 
  Z_c(u,\lambda,\epsilon)=1+\frac{n+2}{3}\, 
  \frac{[1+\lambda\,i'_1(\lambda;m)]u}{2\epsilon}+\Or(u^2)\;. 
\end{equation} 
Finally, $Z_1$ and $P_\lambda$ are given by 
equation~(\ref{eq:Z1Plam}). 
 
In the double integral $J$ we introduce new variables $s$ and $t$ via 
$\hat{P}=\sqrt{t/s}$ and $\hat{p}=s^{-1}\sqrt{1-t^2}$, whereby $J$ 
becomes  
\begin{eqnarray}\fl 
  \label{eq:Jcalc} 
J= \frac{1}{8}\,{\int_0^1}{\rmd} t\,t^{(m-2)/2}  
\,{\left( 1 - t^2 \right) }^{(2 - m - 2\,\epsilon)/4} 
\,{\int_0^\infty}{\rmd} s\, 
  \frac{s^{-2 + \epsilon }\, 
    \left( 1 - c\,s - t\,\lambda  \right) }{ 
    \left( 1 + s\,\kappa_{\bm{p}}  \right) \, 
    \left( 1 + c\,s + t\,\lambda  \right) } 
\nonumber\\ \lo = \frac{{\mathcal N}_{\epsilon,m}}{8}\,  
\left\langle  
{\int_0^\infty}{\rmd} s\, 
  \frac{s^{-2 + \epsilon }\, 
    \left( 1 - c\,s - t\,\lambda  \right) }{ 
    \left( 1 + s\,\kappa_{\bm{p}}  \right) \, 
    \left( 1 + c\,s + t\,\lambda  \right) } 
\right\rangle_{\epsilon,m} 
\;, 
\end{eqnarray} 
where $\langle .\rangle_{\epsilon,m}$ and ${\mathcal N}_{\epsilon,m}$ 
are the average and the normalization factor introduced by 
equations~(\ref{eq:aveps}) and (\ref{eq:normN}), respectively. The 
integral $\int_0^\infty\rmd s$ can be performed analytically and its 
Laurent series coefficients of order $\epsilon^{-1}$ and $\epsilon^0$ 
determined. One thus arrives at 
\begin{eqnarray}\fl 
  \label{eq:chi11Ou} 
  \frac{u\,K_{d-m-1}K_m}{2F_{m,\epsilon}}\,J&=&-\frac{u}{2}\, 
  \bigg\{\left[\epsilon^{-1}-\ln(2\kappa_{\bm{p}})\right] 
  \left[\kappa_{\bm{p}}\,i_1(\lambda,\epsilon;m) 
  +2c\left\langle(1+\lambda  
      t)^{-2}\right\rangle_{\epsilon,m}\right]\nonumber \\ &&\strut  
+2c^2\left\langle\frac{\ln[c/\kappa_{\bm{p}}(1+\lambda t)]}{(1+\lambda 
    t)^2[(1+\lambda 
    t)\kappa_{\bm{p}}-c]}\right\rangle_{0,m}+\Or(\epsilon)\bigg\}\;.  
\end{eqnarray} 
 
We now substitute this result together with the above expressions for 
the renormalization functions $Z_1$, $Z_c$ and $P_\lambda$ into 
equation~(\ref{eq:chi11}) and use the fact that the average in the 
first line of equation (\ref{eq:chi11Ou}) can be written as 
\begin{equation} 
  \label{eq:i2} 
  \left\langle(1+\lambda  
      t)^{-2}\right\rangle_{\epsilon,m}=\case12 
    \left[1+i_1(\lambda,\epsilon;m)+ 
      \lambda\,i_1'(\lambda,\epsilon;m)\right]. 
\end{equation} 
The poles in $\epsilon$ are found to cancel, and one arrives at the 
result (\ref{eq:chi11pt}).

\section*{References} 
\begin{harvard} 
 
\item %
Binder K 1983 {\em Critical behaviour at surfaces\/} {\em in\/} Domb C and 
  Lebowitz J~L, eds., {\em Phase Transitions and Critical Phenomena\/} (London: 
  Academic) vol.~8 pp. 1--144 
 
\item %
Binder K and Frisch H~L 1991 {\em Dynamics of surface enrichment: a theory 
  based on the {K}awasaki spin--exchange model in the presence of a wall\/} 
  {\em Z.\ Phys.\ B\/} {\bf 84} 403 
 
\item %
Bray A~J and Moore M~A 1977 {\em Critical behaviour of semi--infinite 
  systems\/} {\em J.\ Phys.\ A\/} {\bf 10} 1927 
 
\item %
Cardy J~L 1987 {\em Conformal invariance\/} {\em in\/} Domb C and Lebowitz J~L, 
  eds., {\em Phase Transitions and Critical Phenomena\/} (London: Academic) 
  vol.~11 pp. 55--126 
 
\item 
Diehl H~W 1986 {\em Field--theoretical approach to critical behaviour at 
  surfaces\/} {\em in\/} Domb C and Lebowitz J~L, eds., {\em Phase Transitions 
  and Critical Phenomena\/} (London: Academic) vol.~10 pp. 75--267 
 
\item 
Diehl H~W  1997 {\em The theory of boundary critical phenomena\/} {\em Int.\ J.\ Mod.\ 
  Phys.\ B\/} {\bf 11} 3503 ({\it Preprint} cond-mat/9610143) 
 
\item[] 
Diehl H~W  2002 {\em Critical behavior at $m$-axial {L}ifshitz points\/} {\em Acta 
  physica slovaca\/} 52 271 proc.\ of the 5th International Conference 
  ``Renormalization Group 2002'', Tatranska Strba, High Tatra Mountains, 
  Slovakia, March 10--16, 2002 ({\it Preprint} cond-mat/0205284)
 
\item %
Diehl H~W and Dietrich S 1981 {\em Field--theoretical approach to multicritical 
  behavior near free surfaces\/} {\em Phys.\ Rev.\ B\/} {\bf 24} 2878 
 
\item[] 
Diehl H~W and Dietrich S 1983 {\em Multicritical behaviour at surfaces\/} {\em 
  Z.\ Phys.\ B\/} {\bf 50} 117

\item 
Diehl H~W, Gerwinski A and Rutkevich S 2003a {\em Boundary critical behavior at 
  $m$-axial {L}ifshitz points for a boundary plane parallel to the modulation 
  axes\/} {\em Phys.\ Rev.\ B\/} {\bf 68} 224428 ({\it Preprint}  cond-mat/0308483) 
 
\item 
Diehl H~W, Rutkevich S and Gerwinski A 2003b {\em Surface critical behaviour at 
  $m$-axial {L}ifshitz points: continuum models, boundary conditions and 
  two-loop renormalization group results\/} {\em J. Phys. A: Math. Gen.\/} {\bf 36} L243 
 \item 
Diehl H~W and Shpot M 1994 {\em Surface critical behavior in fixed dimensions 
  $d<4$: Nonanalyticity of critical surface enhancement and massive field 
  theory approach\/} {\em Phys.\ Rev.\ Lett.\/} {\bf 73} 3431 
 
\item 
Diehl H~W and Shpot M  1998 {\em Massive field-theory approach to surface critical behavior in 
  three-dimensional systems\/} {\em Nucl. Phys. B\/} {\bf 528} 595 ({\it Preprint} cond-mat/9804083) 
 
\item 
Diehl H~W and Shpot M  2000 {\em Critical behavior at ${m}$-axial {L}ifshitz points: field-theory 
  analysis and ${\epsilon}$-expansion results\/} {\em Phys.\ Rev.\ 
  B\/} {\bf 62} 12 
  338 ({\it Preprint} cond-mat/0006007)
\item 
Diehl H~W, Shpot M and Zia R~K~P 2003c {\em Relevance of space anisotropy in the 
  critical behavior of $m$-axial {L}ifshitz points\/} {\em 
  Phys. Rev. B\/} {\bf 68} 
  224415 ({\it Preprint} cond-mat/0307355) 
 
\item 
Dietrich S and Diehl H~W 1981 {\em Critical behaviour of the energy density in 
  semi--infinite systems\/} {\em Z.\ Phys.\ B\/} {\bf 43} 315 
 
\item 
Drewitz A, Leidl R, Burkhardt T~W and Diehl H~W 1997 {\em Surface critical 
  behavior of binary alloys and antiferromagnets: dependence of the 
  universality class on surface orientation\/} {\em Phys.\ Rev.\ 
  Lett.\/} {\bf 78} 
  1090 ({\it Preprint} cond-mat/9609245) 
 
\item 
Frisch H~L, Kimball J~C and Binder K 2000 {\em Surface critical behaviour near 
  the uniaxial {L}ifshitz point of the axial next-nearest-neighbour {I}sing 
  model\/} {\em J. Phys.: Condens. Matter\/} {\bf 12} 29 
 
\item 
Fr{\"o}hlich J and Pfister C~E 1987 {\em Semi-infinite {I}sing model. 1. 
  Thermodynamic functions and phase diagram in absence of magnetic field\/} 
  {\em Commun. Math. Phys.\/} {\bf 109} 493 
 
\item 
Gel'fand I~M and Shilov G~E 1964 {\em Generalized Functions\/} (New York and 
  London: Academic) vol.~1 pp. 1--423 
 
\item 
Gumbs G 1986 {\em Surface critical behavior near the {L}ifshitz point of a 
  semi-infinite system\/} {\em Phys. Rev. B\/} {\bf 33} 6500 
 
\item 
Hornreich R~M 1980 {\em The {L}ifshitz point: phase diagrams and critical 
  behavior\/} {\em J. Magn. Magn. Mater.\/} {\bf 15--18} 387 
 
\item 
Landau D~P and Binder K 1990 {\em {M}onte {C}arlo study of surface phase 
  transitions in the three-dimensional {I}sing model\/} {\em Phys.\ Rev.\ B\/} 
  {\bf 41} 4633 
 
\item 
Leidl R and Diehl H~W 1998 {\em Surface critical behavior of bcc binary 
  alloys\/} {\em Phys.\ Rev.\ B\/} {\bf 57} 1908 ({\it Preprint} cond-mat/9707345) 
 
\item 
Leidl R, Drewitz A and Diehl H~W 1998 {\em Wetting phenomena in bcc binary 
  alloys\/} {\em Int.\ J.\ Thermophys.\/} {\bf 19} 1219 ({\it Preprint} cond-mat/9704215) 
 
\item 
Pleimling M 2002 {\em Surface critical exponents at an uniaxial {L}ifshitz 
  point\/} {\em Phys. Rev. B\/} {\bf 65} 184406/1 
 
\item 
Pleimling M 2002  2004 {\em Critical phenomena at perfect and non-perfect surfaces\/} {\em J. 
  Phys. A: Math. Gen.\/} {\bf 37} R79 
 
\item 
Pleimling M and Henkel M 2001 {\em Anisotropic scaling and generalized 
  conformal invariance at {L}ifshitz points\/} {\em 
  Phys. Rev. Lett.\/} {\bf 87} 
  125702/1 ({\it Preprint} hep-th/0103194) 
 
\item 
Pfister C~E and Penrose O 1988 {\em Analyticity properties of the surface free 
  energy of the {I}sing model\/} {\em Commun. Math. Phys.\/} {\bf 115} 691 
 
\item 
Ruge C, Dunkelmann S and Wagner F 1992 {\em New method for determination of 
  critical parameters\/} {\em Phys.\ Rev.\ Lett.\/} {\bf 69} 2465 
 
\item 
Ruge C, Dunkelmann S, Wagner F and Wulf J 1993 {\em Study of the 
  three-dimensional {I}sing model on film geometry with the cluster {M}onte 
  {C}arlo method\/} {\em J.\ Stat.\ Phys.\/} {\bf 73} 293 
 
\item 
Schmid F 1993 {\em Surface order in body--centered cubic alloys\/} {\em Z.\ 
  Phys.\ B\/} {\bf 91} 77 
 
\item 
Selke W 1992 {\em Spatially modulated structures in systems with competing 
  interactions\/} {\em in\/} Domb C and Lebowitz J~L, eds., {\em Phase 
  Transitions and Critical Phenomena\/} (London: Academic) vol.~15 
pp. 1--72  
 
\item 
Shapira Y, Becerra C, Oliveira N Jr and Chang T 1981 {\em Phase diagram, 
  susceptibility, and magnetostriction of {MnP}: evidence for a {L}ifshitz 
  point\/} {\em Phys. Rev. B\/} {\bf 24} 2780 
 
\item 
Shpot M and Diehl H~W 2001 {\em Two-loop renormalization-group analysis of 
  critical behavior at $m$-axial {L}ifshitz points\/} {\em 
  Nucl. Phys. B\/} {\bf 612} 
  340 ({\it Preprint} cond-mat/0106105) 
 
\end{harvard} 
 
\end{document}